\documentclass[12pt]{article}
\usepackage{graphicx,amsfonts}
\usepackage{biograph,latexsym}
\usepackage{amssymb,amsmath,multicol}
\usepackage{fullpage}
\usepackage{epstopdf}
\usepackage{color}
\newcommand{\va}{ \mathbf{v} }
\newcommand{\da}{ \mathbf{d} }

\newcommand{\vba}{ \bar{\va} }
\newcommand{\ub}{ \mathbf{u} }
\newcommand{\rb}{ \beta}
\newcommand{\vb}{  \mathbf{v_{\lambda} }  }
\newcommand{\Db}{ \mathbf{D_{\lambda}}  }

\newcommand{\x}{ \mathbf{x} }
\newcommand{\xb}{ \bar{\x} }

\newcommand{\re}{ \mathbb{R} }
\newcommand{\E}{ \mathrm{E} }

\newcommand{\pa}{p(\x,t)}

\begin{document}


\title{Multiscale Analysis of Collective Decision--Making in Swarms: An Advection-Diffusion with Memory  Approach}
\author{M. ~Raghib,
S.A. ~Levin,
I.G.~Kevrekidis}


\maketitle
\begin{abstract}
We propose a (time) multiscale  method for the coarse-grained analysis of self--propelled particle models of swarms comprising a mixture of `na\"{i}ve' and `informed' individuals, used to address questions related to collective motion and collective decision--making in animal groups.  The method is based on projecting the particle configuration onto a single `meta-particle' that consists of the group elongation and the mean group velocity and position.  The collective states of the configuration can be associated with the transient and asymptotic transport properties of the random walk followed by the meta--particle.  These properties can be accurately predicted by an advection-diffusion equation with memory (ADEM) whose parameters are obtained from a mean group velocity time series obtained from a single simulation run of the individual--based model.
\end{abstract}

\paragraph{keywords }continuous time random walks, anomalous transport, collective animal behavior, non-Markovian stochastic processes, self--propelled particle models.



\section{Introduction}
Self-propelled particle models (SPP's) are a class of agent--based simulations that have been used over the last three decades to explore questions related to various kinds of collective motion in animals, including  insect swarms, bird flocks and fish schools 
 \cite{aoki82,reynolds87,huth92,grunbaum94,czirok99,couzin02,parrish02,sumpter06,nabet09}.
In these models, each individual in the (finite) population is represented by a particle that moves with constant speed in two or three-dimensional Euclidean space or a 2-dimensional torus.  All  particles update their  orientations  according to a set of local averages of the current state of the configuration.  These local averages are  simplified representations of individual behaviors that depend on  `social interactions' --avoidance of collisions, attraction, and orientation alignment--  which result in the remarkable property of cohesive collective motion; i.e. the particles move about in space, yet they appear to move as a single object,  resembling the motion of real flocks \cite{aoki82,reynolds87,couzin02}. Errors made by the individuals as  they estimate these quantities are modeled by a random rotation of the output of this averaging procedure.  \\
More recently, SPP models of flocking  have been introduced in the context of collective decision-making to illuminate the question of how groups of agents achieve consensual decisions without the need of a central control \cite{couzin02,conradt03,couzin05,conradt09,nabet09}.  Each of these  decisions can be associated with a variety of collective states, which  typically involve switching between mobile/immobile regimes \cite{kolpas07}, rotation or milling \cite{couzin02},  motion with a directional bias \cite{couzin05}, or a combination of these \cite{moon07}.   A directional bias is relevant when critically important information, for instance  the location of a resource, a predator or a migratory route,  is available only to a fraction of the population \cite{couzin05,moon07}.  \\ \cite{couzin05}  explored this situation using a modified version of earlier models of swarming \cite{aoki82,reynolds87,huth92,couzin02},  where the main innovation consisted of dividing the population into two types.  The first of these, called `na\"ive',  follow only the social rules mentioned earlier (avoidance, attraction and alignment).  The second kind, dubbed `informed', also obey the social interactions of the na\"{i}ve individuals, but weigh the social output with an orientation bias along a single `preferred' direction, which in this study is identical for all informed individuals.  This orientation bias can be regarded as a simple representation of access to privileged information. Collective decision--making is understood in this context in terms of  the ability of the informed sub--population  to transfer their orientation bias to the whole group while simultaneously preserving group cohesion. \\

Despite the recent explosion of SPP models in the literature,  our understanding of these systems still remains limited.  Central challenges are related to our ability  to characterize efficiently and meaningfully the dynamics of each collective state, and critically, their dependence on the parameters of the individual--level model.
We identify three distinct approaches to address this problem; namely Monte-Carlo simulation, continuum models, and `hybrid' multi-scale approaches. \\  
The first (the Lagrangian approach) is mainly computational and consists of moving with each individual particle.  Macroscopic summary statistics describing the various collective states are obtained  from  averages based on a large number of independent simulation runs, or a single time series when ergodicity is a reasonable assumption.  These average quantities usually include the mean group velocity \cite{czirok99}, the mean angular momentum \cite{couzin02}, mean switching times between mobile/immobile states  \cite{kolpas07} or the `accuracy' of the decision-making process \cite{couzin05,merrifield06}. Other state variables of interest link collective states to geometrical properties of the flock, like the group elongation \cite{couzin05} or its aspect ratio \cite{cavagna06}. \\
The second  method (the Eulerian approach)
focuses on continuum models for the density and velocity fields.  It has the advantage that in some cases analytical results linking the microscopic to the macroscopic can be rigorously derived. In addition to this, the numerical solution of the model for large or small population densities has the same computational cost, and the mechanisms that generate the collective patterns can often be clearly distinguished in the various terms in the model, which provides some degree of parsimony that approaches based solely on Monte--Carlo simulations cannot emulate.  Continuum approximations have been used to approximate discrete SPP models mainly to study collective motion that is not cohesive \cite{czirok99}; i.e. the population lives in a spatial arena with periodic or reflecting boundaries but does not form a single distinct group.  Instead, particles move about freely forming and dissolving groups of various sizes (i.e. fission--fusion dynamics), and collective motion is detected as a non--vanishing population average of the velocity.  These continuum models are obtained through heuristic reasoning based on careful observation of system symmetries, or the invocation of conservation laws \cite{mogilner99,toner05,topaz06,eftimie07}. \\
Although substantial progress has been made with Eulerian (continuum) approaches, particularly for swarming microbial populations  
\cite{aranson07,sokolov07}, there are still a number of issues that preclude their widespread use.  First, the use of heuristics does not clarify the dependence of the macroscopic  parameters on the individual--level model.  Although  some continuum models have recently been derived formally from the individual--based model via a limiting process (usually large population size), the theoretical progress is made at the expense of great  simplifications which restrict strongly their biological relevance.  For instance \cite{bertin06} and  \cite{chuang07} each derived continuum models in the limit of large population sizes, but restricted the individual--level interactions to a single type of social  interaction, specified via a potential function  \cite{chuang07},  or a velocity average \cite{bertin06}.  Second, they usually require very large population sizes in order to be meaningful, which is problematic for models of flocking in  groups involving tens or perhaps hundreds of individuals. In this situation the finiteness of the population size plays a fundamental role in observed transport properties  (e.g. the group tends to move more slowly as the population size increases) \cite{couzin02,buhl06, cavagna06,grunbaum08,sumpter08,ward08}. \\
The third is the hybrid multiscale approach, which attempts to bridge Monte--Carlo simulations and continuum models.  It is based on assuming the existence of a continuum model for some relevant coarse--grained state variable or `reduction coordinate'; for instance Non-linear Advection--Diffusion Equations (NADE) with density--dependent coefficients \cite{grunbaum08}, or  Fokker--Planck equations with a non-linear potential \cite{kev03,rad05,moon07,kolpas07,coifman08,yates09}, which serves as a model template. The unknown fluxes and coefficients in the macroscopic template are \emph{estimated} from a computational experiment, which usually consists of a single --and relatively short-- simulation run of the microscopic model.  These estimated quantities are substituted into the unknown terms in the macroscopic model, which is then analyzed by means of the appropriate suite of classical continuum methods, numerical or analytical.\\
 
In this study, we use this latter approach to explore the ability of  Continuous Time Random Walks (CTRW) \cite{montroll73,kenkre77,cortis04,berkowitz06},  and its associated continuum counterpart, the Advection--Diffusion Equation with Memory (ADEM) --also known as the Generalized Master Equation (GME)-- as a model template for the coarse--grained dynamics of cohesive collective motion and collective decision--making in self--propelled particle models of swarms comprising a mixture of individuals that have preferential access to critical information --the `informed' type-- and those who do not ( `na\"{i}ve').  The ADEM generalizes the classical advection--diffusion equation to a non--local--in--time transport model via the introduction of a `memory', a time weighting function proportional to the  particle's two--time velocity autocorrelation function.  The ADEM is a useful model of anomalous transport that arises when the underlying random walk possesses a wide distribution of transition rates \cite{kenkre77,kenkre09,cortis04,metzler04,berkowitz06}.  The multiscale method we propose is based on coarse--graining the full SPP configuration into a single `meta--particle', that consists of the group elongation (as a measure that the group remains cohesive) and the mean group velocity and position.  The various types of collective states displayed by the group can then be related to the transport properties of the meta--particle's random walk, under the assumption that the pdf of the transition density for the meta-particle's position follows an ADEM.  \\
We illustrate the method for the case of a 2--dimensional SPP model introduced earlier by  \cite{couzin05} for a single informed direction, but the approach is quite general in the sense that it can be applied to any individual--based model of movement for which the biologically meaningful coarse variables are the mean group position and velocity, and that the effective distribution of jump lengths at each transition event has finite moments of all orders.  The multiscale approach for collective motion based on the ADEM complements \emph{local}--in--time multiscale approaches for a similar class of individual--level models explored earlier \cite{kolpas07,grunbaum08,yates09}.  For instance, the ADEM can predict correctly the transport properties even when the individual--level model has a strong alignment rule, which is precisely the main limitation of the otherwise successful method based on non-linear advection--diffusion equations \cite{grunbaum08}.  This results from temporal correlations in velocity fluctuations induced by the alignment rule that persist over macroscopically relevant time scales, a property that can not be captured by local--in--time  Markovian models, but can be dealt with via the introduction of a memory term. \\
CTRW theory  generalizes the classical Random Walk (RW)  as a  microscopic model better suited for problems in anomalous  transport, which is usually detected when the mean squared displacement (msd)  does not scale linearly with time over a wide range of time scales.  The anomalous properties can frequently be attributed to the presence of a wide distribution of transition rates (or also in the jump lengths), which leads to persistent temporal correlations in velocity fluctuations.  It is the presence of time correlations in velocity that ultimately leads to anomalous transport
\cite{montroll73,kenkre77,klafter80,metzler04}. The variability in transition rates can be attributed in real systems to spatial disorder in the medium, as is the case in tracer transport in porous media \cite{berkowitz06}.  The presence of spatial disorder in the medium  creates localized structures that can trap the particle for long periods of time, or force it to move ballistically by confining its motion along a corridor.  The resulting particle motion consists of alternating bursts of ballistic motion, apparent brownian motion, and a stagnant phase where the particle moves very slowly, if at all.   This resembles the dynamics of the group meta--particle in SPP models of swarms, which typically consists of bursts of alignment in the particle orientations that lead to advective flights at the group level (the slip phase),  alternating with regimes of slow motion when the particles lose their alignment and the mean group velocity drops sharply (the stick phase).  The power of CTRW \cite{cortis04,berkowitz06} and effective medium theories of random motion in disordered media \cite{kenkre09}, lies in that the spatial inhomogeneities in the medium responsible for the anomalies in transport properties are not modeled explicitly.  Instead, their effect is summarized \emph{statistically} in terms of the effective distributions of jump lengths and waiting times that define the random walk.  The key innovation of CTRW theory is that the random walk does not proceed  by fixed spatial and temporal increments, but these  become instead random variables, defined by two probability densities, which are usually assumed independent in applications. The first is the distribution of jumps in space $\lambda(\xi)$, which prescribes the length of the jumps between locations at each transition event. The second is a clock that regulates the times elapsed between transitions, known as the distribution of  waiting times $\psi(\tau)$.  A thorough discussion of modern CTRW theory and  its role in models of anomalous transport can be found in a recent review by \cite{metzler04}.  
 \\
It can be shown \cite{zwanzig65,montroll73,kenkre77,metzler04,berkowitz06} that when the distribution of jump lengths can be expanded in a Taylor series and the distribution of waiting times is an arbitrary probability density function, the transition probability density $p(\x,t|\mathbf{0},0)$ for  finding a particle around position $\x \in \re^2$ at time $t$ given that it started at the origin at time zero, obeys a modified version of the advection--diffusion equation that is non-local in time,  known as the Advection-Diffusion Equation with Memory (ADEM) 
\begin{eqnarray}
 \label{eq:ADEM1}
 \frac{\partial \pa}{\partial t}&=&-\int_{0}^{t}\,M(t-s)\,\left[ \vb \cdot \nabla p(\x,s) - \Db\,  
: \,  \nabla \, \nabla p(\x,s)\right] \, ds \\  
\nonumber p(\x,0^{+}) &=& \delta(\x), ~~\x \in \re^{2},~~t\in \re^{+}
 \end{eqnarray}
where  $\vb$ is the effective drift vector, $\Db$ the diffusivity tensor,  and the colon operator is the inner tensor product
\[
\mathbf{A}:\mathbf{B} = \mathrm{Trace} \{\mathbf{B}^{T} \cdot \mathbf{A} \} =\sum_{i,j} A_{ij}B_{ij}.
\]
The transport coefficients $\vb$ and $\Db$ are determined respectively by the ratio of the first two moments of the jump distribution to the mean of the waiting time distribution, or the median when $\psi(t)$ does not have a finite mean \cite{berkowitz06}.  The memory function $M(t)$ has two equivalent interpretations; it is closely related to the distribution of waiting times \cite{cortis04}, but it can also be shown to be proportional to the velocity time auto--correlation function of the moving particle \cite{haus87,kenkre03,west97}.
\[\E \left[v_{1}(0)\,v_{1}(\tau)\right] = 2D_{1} M(\tau),
\]  
where $v_{1}$ is the velocity along the $x_{1}$ direction, $D_{1}$ is the diffusivity along $x_{1}$ and $M(t)$ is the memory kernel that prescribes the decay of correlations (see Section \ref{sec:multiscale} for additional details). This model  constitutes the basis for effective medium theories of anomalous transport in disordered media, where the spatial disorder in the medium is replaced by an ordered model with  memory of the form  (\ref{eq:ADEM1}) \cite{haus87,kubo91,kenkre03,berkowitz06,kenkre09}.  \\
The stochastic dynamics of the meta--particle associated with  the SPP model of flocking explored here has a striking resemblance to that which motivated the development of the theory for anomalous transport in heterogenous media based on the CTRW and the ADEM.  In SPP models, the wide range of variability in transition rates cannot be attributed to  spatial disorder in the medium, but  arises instead from stochasticity in the alternating (slip/stick) types of collective behavior. Even though the source of variability is quite different, this does not seem to matter provided that transport can be modeled in terms of \emph{effective} distributions of jump lengths and waiting times and their associated ADEM.
 Our goal is to exploit this analogy to propose an ADEM as a continuum `model template' for the dynamics of the position pdf of a swarm centroid.  The functional form of the memory and the transport parameters in the ADEM template can be estimated from a single mean group velocity time series obtained from a  simulation run of the SPP model.  The resulting fitted model can be then used to explore the dependence of the collective behaviors on the parameters that determine the individual--level model, particularly the strength of the bias of the informed sub--population, the total population size, and the proportion of informed individuals. \\

In the ADEM approach, the memory is the fundamental object that encodes all the transport coefficients, the various transport regimes and their characteristic timescales \cite{kubo57,kenkre03}.  When the spatial distribution of the disorder is known \cite{kenkre09} or the Hamiltonian of the microscopic model \cite{kenkre03},  it is possible to derive the memory in (\ref{eq:ADEM1}) from the microscopic dynamics.  In general, one has to resort to simulations or experiments and subsequent function fittings, in order to obtain the velocity auto--correlation function.  The non-linearities involved in the definition of the SPP seem to preclude the derivation of the velocity time auto--correlation function  rigorously from the microscopic swarm model.   We find from simulations that the memory kernel along the informed direction for SPP models can be very well fitted by two closely related functions. The first corresponds to a Gamma density,
\begin{equation}
\label{eq:gamma1}
M(t)=\frac{\tau_a^{\gamma-1}}{\Gamma(1-\gamma)}t^{-\gamma}\exp(-t/\tau_a),
\end{equation}
which works well in swarms where there are no informed individuals present, but also when the proportion of informed individuals is small (and relatively low values of the coupling strength).  The initial power law decay in (\ref{eq:gamma1}) leads to a sub--ballistic, super--diffusive transient detectable in the mean--squared displacement.  This power law behavior has an exponential truncation at a characteristic time scale $\tau_a$ that establishes the onset of  the asymptotic regime, which is dominated by diffusion in swarms with no informed individuals and a mixture of diffusion and  advection (with constant drift) for groups that include informed individuals.  We also find that the diffusion coefficient decreases with group size, and the time scale ($\tau_a$) that determines the onset of the asymptotic regime increases with group size.\\
The second, `richer' situation,  arises in informed swarms for high values of the bias along the informed direction, where the early time super-diffusive transient is followed by a regime where correlations oscillate before reaching the asymptotic state, which is also classical advection--diffusion.  This additional regime requires a modification of the memory kernel (\ref{eq:gamma1}) in order to capture these oscillations.  We find that a Mittag--Leffler function $E_{\alpha,\beta}(z)$ with an exponential truncation \cite{podlubny99,west03},
 \begin{equation}
 \label{eq:mlf1}
 M(t) =\frac{\tau_s + \tau_a^\alpha}{\tau_s \tau_a^\beta}\, t^{\beta-1}\,E_{\alpha,\beta}\left[-\left(t/\tau_s\right)^{\alpha}\right]\,\exp(-t/\tau_a),
 \end{equation}
provides an excellent fit in this regime, at the cost of introducing two additional parameters (the exponent $\beta$ and the time scale $\tau_{s}$).  We used these estimates together with the ADEM model in order to predict the behavior of the mean squared displacement (msd), i.e. the second moment of the mean group position, which  can be used to characterize the various types of collective behaviors and their characteristic time scales in terms of their effect on the meta--particle's transport properties. The functional forms themselves do not seem to change with group size, but only the parameters do.\\ %
 For the region of parameters where the group remains cohesive, we observed that  there are two types of collective behavior that are shared by both na\"{i}ve (no informed individuals present) and informed groups.  First,  there is  an anomalous super--diffusive transient at early times (the scaling exponent in the mean squared displacement lies between one and two)  due to the prevalence of slip/stick dynamics over that domain of time scales. Asymptotically, the msd scales linearly with time for na\"{i}ve groups (diffusion--dominated), but shows a sharp transition to quadratic scaling (advection--dominated) for informed ones along the informed direction, which indicates that on average, informed swarms diffuse, but also move with constant velocity over the longer time scales.   This transition from linear to quadratic scaling allows the detection of the time scale at which the informed sub-population manages to transfer its orientation bias to the whole group;  this time scale, or time to consensus, is a natural measure of the efficiency of the decision--making process. The magnitude of the drift, which depends on the degree of polarization of the particle orientations along the informed direction,  is a straightforward macroscopic parameter for the degree of consensus.  We also note that  as the group size gets larger, the drift gets smaller for the same proportion of informed individuals and informed bias strength. Finally, the diffusion coefficient along the informed direction can be interpreted as a measure of the precision of the collective decision--making process --since it is a measure of the spread of an ensemble of swarm meta--particles-- when compared with that of na\"{i}ve configurations.\\
The resulting ADEM fitted from swarm simulation time-series is self-consistent in the sense that transport parameters estimated from the memory via a Kubo--Green relationship \cite{kubo57,green60} coincide with those estimated from the moments of the jump and waiting time pdf's of the associated CTRW for the three   group sizes explored ($N=10,50,100$), proportions of informed individuals, and strength of the bias along the preferred direction.  We also discuss the phase diagrams for the transport coefficients estimated from this method, where we notice velocity--precision trade--offs: as the total group size gets larger, the decision--making becomes more precise at the expense of a slower mean group velocity.   We also note that the time scale to consensus is invariant  with respect to group size, and depends only on the proportion of informed individuals and the strength of the coupling along the informed direction.\\

The paper is organized as follows:  Section \ref{sec:spp} introduces a slightly modified version of the SPP model with informed individuals of  \cite{couzin05}, where we removed the  constraint on the maximum turning angle that an individual can make during a time step.  We then define the set of coarse--grained variables of interest, namely  the group elongation, the mean group position, and the mean group velocity which we called the meta--particle.  Simulation results are also shown, focusing on the mean squared displacement of the meta--particle as well as kernel density estimates of the  probabilities of mean group speeds and orientations, finalizing with  group elongation time series that detect when the group splits appart.  These results are later used to define macroscopic measures of collective motion and collective decision--making in terms of the transport regimes  shown in the msd.   Section
\ref{sec:ctrw} briefly reviews known results from the theory of continuous time random walks (CTRW) \cite{montroll73,kenkre77}, and its relationship to the advection-diffusion equation with memoy (ADEM) \cite{berkowitz06} that we use later as the macroscopic transport model for the transition density of the mean group position.  Section \ref{sec:multiscale} assumes that the random walk followed by the group meta-particle evolves according to a CTRW, and discusses the procedure used to estimate the memory and the transport coefficients of the associated ADEM, from a single velocity time series obtained from  a run of the individual-based model. We compare mean squared displacements obtained from ensemble averages over simulation runs with those predicted by the fitted ADEM for which show analytical results for the time to consensus.  The method is used to carry out a systematic exploration of the dependence of the macroscopic parameters --the diffusivity, the drift and the time to consensus-- on the microscopic ones of immediate biological relevance; namely the relative proportion of  informed individuals, the coupling strength, and the total population size.   Some final remarks are presented in Section \ref{sec:discussion}.\\

\section{Self-propelled particle model (SPP)  with informed individuals}
 \label{sec:spp}
\begin{figure}
\begin{center}
\includegraphics[width=4in]{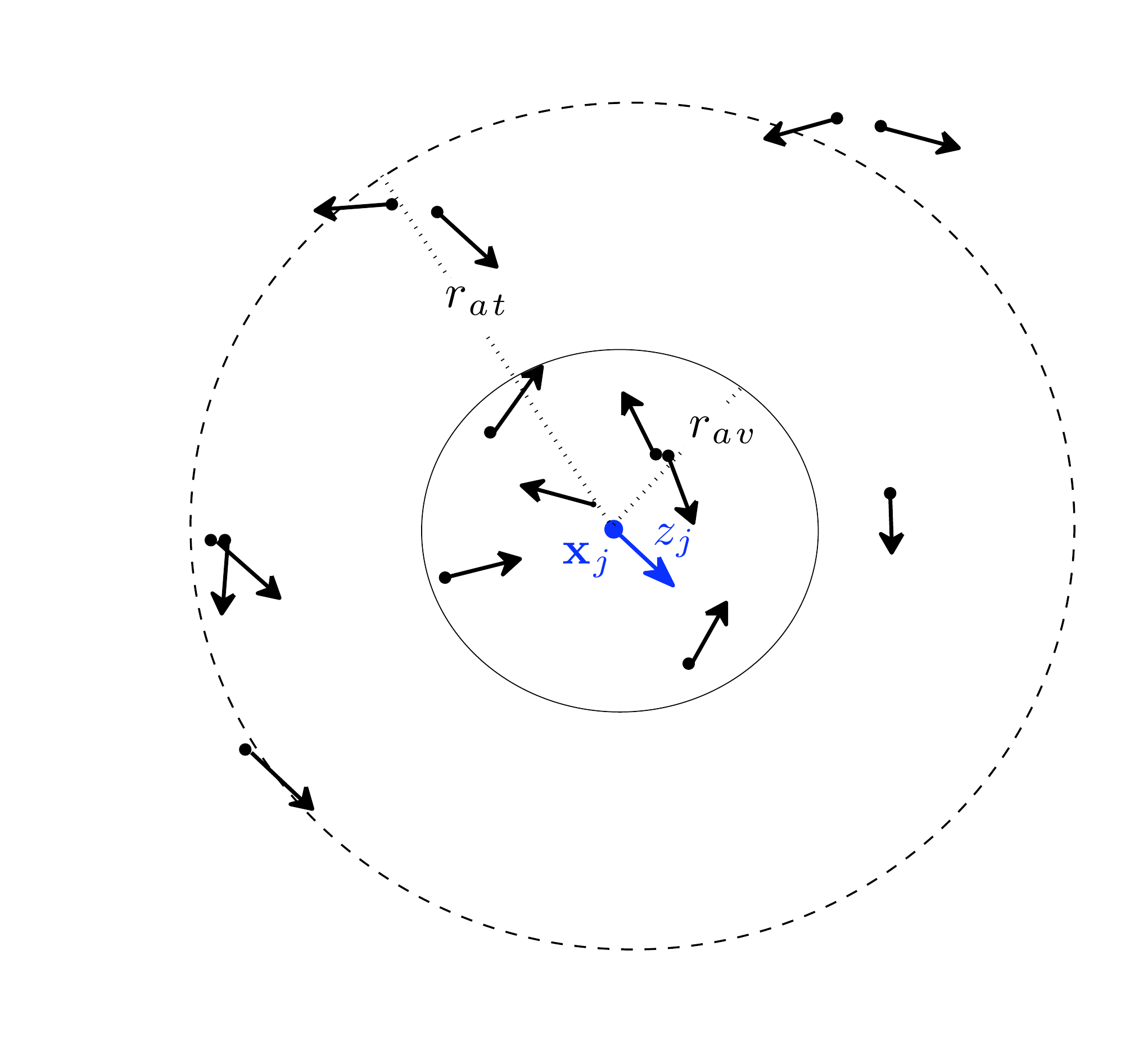}
 \caption{Interaction zones for a focal individual (blue) $\x_j$ (dot) with velocity $z_j$ (arrow). The dots and the arrows represent other particles in the configuration (black). The region of avoidance is the interior of the circle of radius $r_{av}$.  The particles contributing to the region of alignment and attraction lie within the annulus of external radius $r_{at}$ and internal radius $r_{av}$.}
 \label{fig:interaction_zones}
 \end{center}
\end{figure}

Consider a population of $j=1,\ldots,N$ particles  with  positions $\x_j (t)$ in 2-dimensional Euclidean space.  Each particle $j$  moves  with constant speed $s$  along its orientation angle $\theta_j(t)$  in $[-\pi,\pi)$.  We summarize this information as the (complex)  particle velocity
\[z_j(t)= s\,e^{i \,\theta_j(t)}.
\] 
The state of the  population at (discrete) time $t$  is represented by the configuration $\Phi_t(A)$
 \begin{equation}
   \label{eq:config}
  \Phi_t (A) =\left\{\, [\,\x_j(t),z_j(t) \,] \right\}, 
 \end{equation}
where  $A$ is the region of observation.  At each tick of the clock, the positions and orientations of each particle are updated according to,
 \begin{eqnarray}
   \label{eq:updatePos}
   \x_j(t+\Delta t) &=& \x_j(t)+s\,\left(\begin{array}{c} \cos[\theta_j(t)\,] \\ 
                                                  \sin[\theta_j(t)\,]  
                                   \end{array}
                              \right) \Delta t \\                             
   \nonumber
   \theta_j(t+\Delta t) &=& \langle  \,  \Phi_t (D_j)  \,\rangle \,\exp(i \, \Delta Q)   
 \end{eqnarray}
 where $\Delta t$ is the time increment and $\langle  \Phi_t (D_j) \rangle $ is a local average of the configuration restricted to an interaction region $D_j$ centered around the $j$-th particle. The details of the averaging procedure are described in the collective motion rule below ( Figure \ref{fig:interaction_zones}).  Errors made by the individuals in their estimates of the local state of the configuration are modeled by rotating the updated orientation obtained from the local average by a random angle $\Delta Q$, drawn from the wrapped Gaussian on the unit circle $\mathcal{N}_{w} (0,\sigma^{2}\,\Delta t)$ with mean zero and variance $\sigma^2 \Delta t$. \\
 The local average in (\ref{eq:updatePos}) comprises two groups of  rules.  The first is based on the classical social interactions for collective motion, with parameters restricted to the domain in which the full configuration moves cohesively as a single object
 \cite{aoki82,reynolds87,czirok99,flierl99,couzin02}. The second is a steering rule proposed by Couzin \emph{et al} \cite{couzin05},  that attempts to lead the motion of the group along a preferred orientation $\rb$.  This additional rule is followed only by a sub-population of `informed individuals'.   Whereas  individuals that are not informed (called `na\"{i}ve') update their orientations exclusively from the output of the social rules, informed individuals update their orientations according to a weighted average of the social interactions with  the preferred direction. The weight of the bias along the preferred direction relative to the social rules is given by a `coupling constant' $\omega$,  which is interpreted as a simple parameterization of an `internal state' of the informed individual (e.g. starvation, detection of a predator or a resource).   Collective decision--making is  then understood in terms of the ability of the informed sub-population to transfer their orientation bias to the whole group.  \\
 
The three social interactions are: 1) avoidance of collisions, 2) attraction (centering), and 3) alignment (polarization).    Whereas the collective motion interactions are followed by all  $N$ particles, the steering rule is followed only by the informed sub-population of $N_{\rb} \leq N$ particles, whose indices $J_\beta=\{j_1,\ldots, j_{N_\rb}\}$ are chosen uniformly  from the set of indices of all the particles in the configuration $J_\Phi=\{1,\ldots,N\}$.  Both the number of informed particles as well as their indices remain fixed for all times once chosen at time zero. The particles that are not in the informed sub-group are called called `na\"{i}ve'.   Following Couzin \emph{et al} 
 \cite{couzin05} we have
  
 \begin{enumerate}
  \item{\textbf{Collective motion rule}}
  
    \begin{enumerate}
   
    \item{Avoidance of collisions} \\
     We define the neighborhood of avoidance of  
     the $j$-th particle $Av_j=B(r_{av},\x_{j}(t))$  as the circular domain of radius $r_{av}$ 
     centered at $\x_j(t)$ (see Figure \ref{fig:interaction_zones}).  
     If the configuration restricted to the window $Av_j$ is not empty, the avoidance rule takes 
     precedence over the other interactions.  The avoidance rule prevents   
     collisions by pointing the focal particle in the opposite direction of the 
     centroid of the locations of the particles found within $Av_j$, relative to the location 
     of the focal particle $\x_j(t)$. The number of neighbors of the $j$-th  
     particle in $\Phi_t(Av_j)$ is
     \begin{equation}
      \nonumber
     N_{Av_j} =\sum_{k \neq j}^N I_{Av_{j}}\left(\x_k(t)\right)
     \end{equation}
      where the focal individual $j$ is excluded from the count and $I_{B}(\x)$ stands 
     for the indicator function of some 2-D domain $B$,
     \begin{eqnarray}
        \label{eq:indicatorF}
         I_{B}(\x)=\left\{\begin{array}{cc} 1, & \mbox{ if } \x \in B \\
                                          0 & \mbox{ otherwise}.
                        \end{array}\right.
     \end{eqnarray}
     The vector pointing in the direction opposite to the centroid of the particles in $A_{v_j}$ is
     \begin{eqnarray}
        \label{eq:avoidanceRule1}
       \da_j(t+\Delta t) &=& -\frac{1}{N_{Av_j}}\sum_{k \neq j}^{N} I_{Av_{j}}\left(\x_k(t)\right)
     \left[\, \x_k(t)-\x_j(t)\,\right],
     \end{eqnarray}
 The updated orientation due to avoidance is
     \begin{eqnarray}
       \label{eq:avoidanceRule3}
       \theta_j(t+\Delta t) &=&\arg(\da_j)+ \Delta Q,
     \end{eqnarray}
     where $\Delta Q$ is a random angle drawn 
     from  $\mathcal{N}_w(0,\sigma^2\,\Delta t)$.\\
     
    \item{Attraction and Alignment}\\  
     If the configuration restricted to $Av_j$ is empty, 
     we proceed to evaluate 
     the alignment and attraction updating rules.  The neighborhood of attraction/alignment of the $j$-th particle  
     is $At_{j}=B(r_
     {at},\x_{j}(t)\,)$, where  $B(r_{at},\x_{j}(t)\,)$ is the circular domain of radius $r_{at}$ 
     centered at $\x_{j}(t)$. 
     The social interaction in this case is given by the normalized vector 
     sum over the positions (which determines the local attraction vector),  and the velocities,   
     (which dictates the local alignment vector) of the neighbors.  
     The number of neighbors in $At_j$ is
     \[
       N_{At_j} =\sum_{k=1}^N I_{At_{j}}\left(\x_k(t)\right), 
     \]
     The contribution due to attraction is given by the vector $\da_j^\xi$ pointing in the 
     direction of the centroid of the positions of the neighbors relative to the focal individual
     \begin{eqnarray}
      \label{eq:attractionRule1}
      \da_j^\xi(t+\Delta t)  =  \frac{1}{N_{At_j}}\sum_{k=1}^{N} I_{At_{j}}\left(\x_k(t)\right)\,
      [\,\x_k(t)-
         \x_j(t)\,] 
      \end{eqnarray}
      and the contribution due to the  alignment  behavior $\da_j^\theta$ comes from 
      the average 
      orientation of all the particles in $At_{j}$
      \begin{eqnarray}
      \label{eq:attractionRule2}
      \da_j^\theta(t+\Delta t) &=& \sum_{k=1}^{N} I_{At_{j}}(\x_k(t))\,
      z_k(t) .
     \end{eqnarray}
     The total contribution of the social rules is given by the vector sum of the normalized vectors associated with the attraction 
  (\ref{eq:attractionRule1}) and alignment contributions (\ref
     {eq:attractionRule2}),
      \begin{eqnarray}
       \label{eq:attractionRule3}
       \da_j(t+\Delta t) &=& \frac{ \da_j^\xi(t+\Delta t)} { \| \da_j^\xi(t+\Delta t)\| }
                       + \frac{ \da_j^\theta(t+\Delta t)}{\|   \da_j^\theta(t+
                         \Delta t)\|}.
      \end{eqnarray}
 These two contributions are equally weighted in (\ref{eq:attractionRule3}) but  could be generalized so as to have different weights.  In what follows we explore the former, mainly to explore the potential of the ADEM to predict the macroscopic dynamics in the presence of strong alignment, which has been shown to be problematic for Markovian models \cite{grunbaum08}.  The updated orientation is given by the argument of the social interactions $\da_{j}$  
     after rotating it by a small random angle $\Delta Q$ drawn as well from the wrapped Gaussian $
     \mathcal{N}_{w}$
     \begin{eqnarray}
       \label{eq:attractionRule5}
       \theta_j(t+\Delta t) &=& \arg(\da_j) + \Delta Q.
     \end{eqnarray}
   \end{enumerate}
   \item{\textbf{Steering rule for the informed sub-population}}\\
    If the index of the focal particle is in the list of informed indices $J_\rb$, the updated direction   
    is given by a compromise between the output of the social rules (\ref
    {eq:attractionRule5}) and the informed individual's preference to move along the informed direction $
    \rb$.  This is given by the weighted vector average of these two contributions ( \ref{fig:informed_rule})
    \begin{eqnarray}
       \label{eq:informedRule1}
        \da^{\ast}_j(t+\Delta t) = \mathbf{u}_j(t+\Delta t)
       + 
       \omega \,  \hat{\mathbf{b}},
    \end{eqnarray} 
where $\omega$ is a weighting constant, $\mathbf{u}_{j}$ is the unit orientation vector arising from the social rules (\ref{eq:attractionRule5}) and $\hat{\mathbf{b}}$ is the unit vector associated with the preferred orientation $\rb$ and $\hat{\da}_j$. The updated orientation is
\begin{equation}
  \label{eq:infInd}
  \theta_j(t+\Delta t)=\arg[ \da^{\ast}_j].
\end{equation}  
\begin{figure}
\begin{center}
 \includegraphics[width=5in]{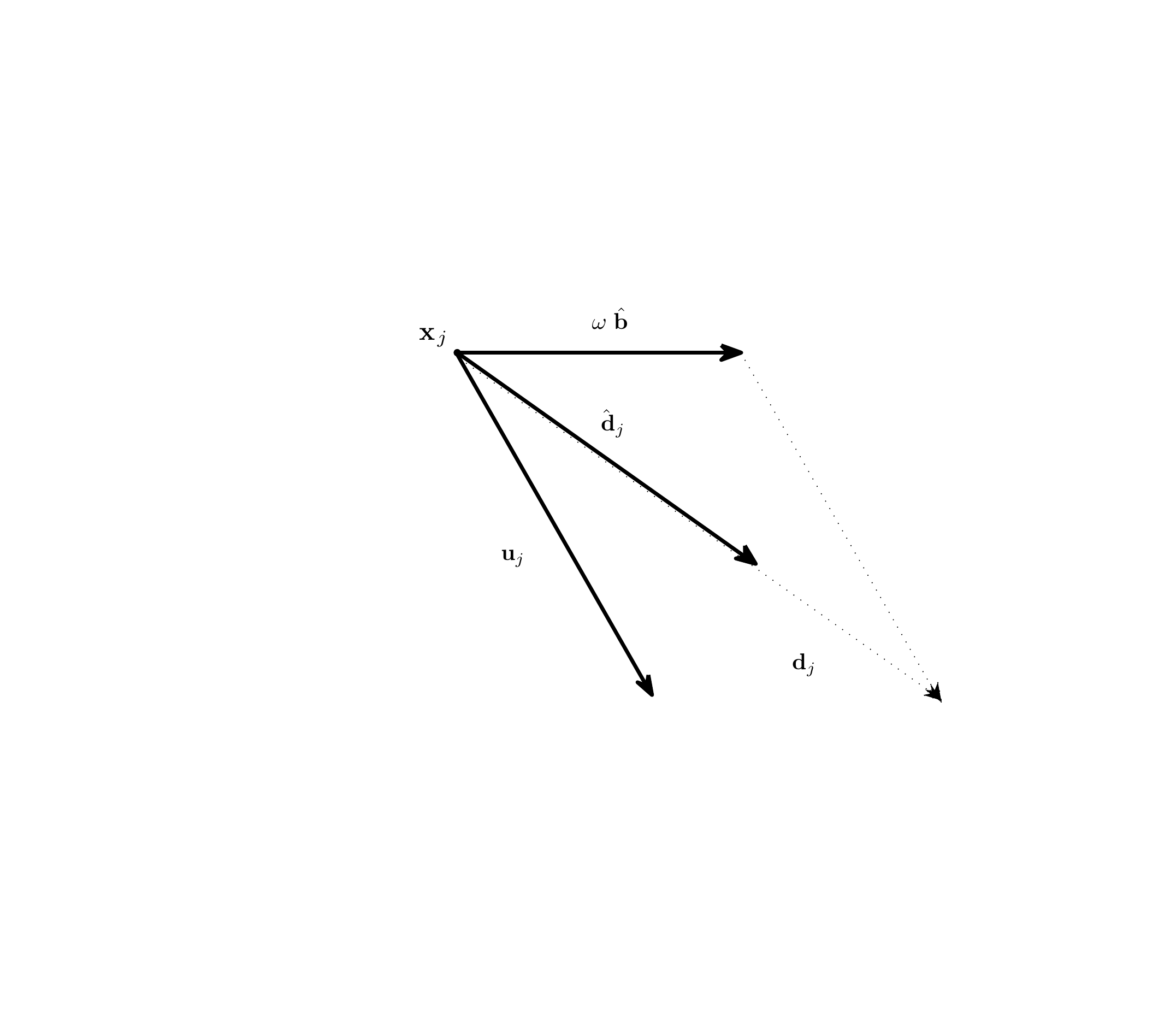}
 \caption{Updating rule for an informed particle.  The updated direction corresponds to the normalized vector sum $\hat{\mathbf{d}}_j$ of the preferential direction vector $\hat{\mathbf{b}}=(\cos(\rb),\sin(\rb))$ rescaled by a factor $\omega$, with the unit vector pointing in the direction of the output of the social rules  $\ub_j=(\cos(\theta_j),\sin(\theta_j))$.}
 \label{fig:informed_rule}
 \end{center}
\end{figure}
Once all the particle's orientations are computed according to these social rules, the positions are updated according to (\ref{eq:updatePos}). A summary of the parameters in the SPP model, together with the values used for the simulations are shown in Table \ref{table:param}.\\
\end{enumerate}
\begin{table}[htdp]
\caption{SPP model parameters}
\begin{center}
\begin{tabular}{|l|c|c|c|}
\hline
Parameter    & symbol & value & units \\ [0.5ex]
\hline
Radius of avoidance                 & $r_{av}$ & 1.0 & m. \\
\hline
Region of avoidance                 & $Av$ & -- & m${}^2$ \\
\hline
Radius of attraction                 & $r_{at}$ & 5.0 &m. \\
\hline
Region of attraction/alignment                 & $At$ &  -- & m$^2$ \\
\hline
Particle speed                 & $s$ & 1.0 & m/sec. \\
\hline
Perception error          & $\sigma$ & 0.1&  radians \\
\hline
Time step          & $\Delta t$ & 0.1& seconds \\
\hline
Total population size          & $N$ & 10 -- 100 & individuals \\
\hline 
Informed population size   & $N_\rb$ & 0 --100 & individuals \\
\hline
Coupling constant    &  $\omega$ & 0.0 -- 0.6 & dimensionless \\ 
\hline
Informed orientation angle   & $\rb$ & 0.0 & radians  \\ [0.5ex]
\hline
\end{tabular}
\end{center}
\label{table:param}
\end{table}

\subsection{Simulation results}\label{ssec:simulationresults}
Our coarse--grained analysis of the individual-based model consists of projecting the full configuration (\ref{eq:config}) onto a set of summary statistics that we dubbed the `meta--particle'. We associate the stochastic properties of the meta--particle random walk to the various collective states of the full configuration.  For collective decision-making, we found that a useful projection consists of three state variables,  the group elongation $\Lambda(t)$, the mean group velocity $\vba(t)$, and the mean group position $\xb(t)$. The  introduction of the group elongation is necessary in order to detect situations where the informed individuals leave the main group, something that occurs at high values of the coupling constant.  In this situation collective decision--making is not consensual, since the informed individuals fail to lead the complete group along the informed orientation. \\

\subsubsection{Projections of the configuration}
 The elongation is defined as the maximum of the set of two--point distances among all the positions of the particles in the configuration,
\begin{equation}
 \label{eq:groupElongation}
\Lambda(t):=\max \left\{ 
              \, \|\x_j(t)-\x_k(t) \| 
                    \left|\frac{}{}\right.
                   \forall \, j,k \in J_\Phi 
                             \right\}.
\end{equation}
We restrict our study to values of the coupling constant $\omega$ that preserve cohesive collective motion, in the sense that  the whole configuration moves as a single entity \cite{reynolds87,couzin02}. This is tantamount to requiring $\Lambda(t)$ to have a constant upper bound $C$, 
\begin{equation}
 \label{eq:cohesiveMotion}
 \Lambda(t) < C  << \infty.
\end{equation}
If the property (\ref{eq:cohesiveMotion}) is preserved,  measures for consensual collective motion and decision--making  can be developed in terms of the stochastic properties of the two other projections of the configuration which together with the elongation, define the configuration `meta-particle' $\varphi_t$ 
\begin{equation}
\label{eq:metaparticle}
\varphi_t =\{\,\Lambda(t),\, \xb(t),\, \vba (t) \, \},
\end{equation} 
where the second element in the triplet  is the mean group position, or configuration centroid $\xb(t)$
\begin{figure}
 \begin{center}
 \includegraphics[width=6in]{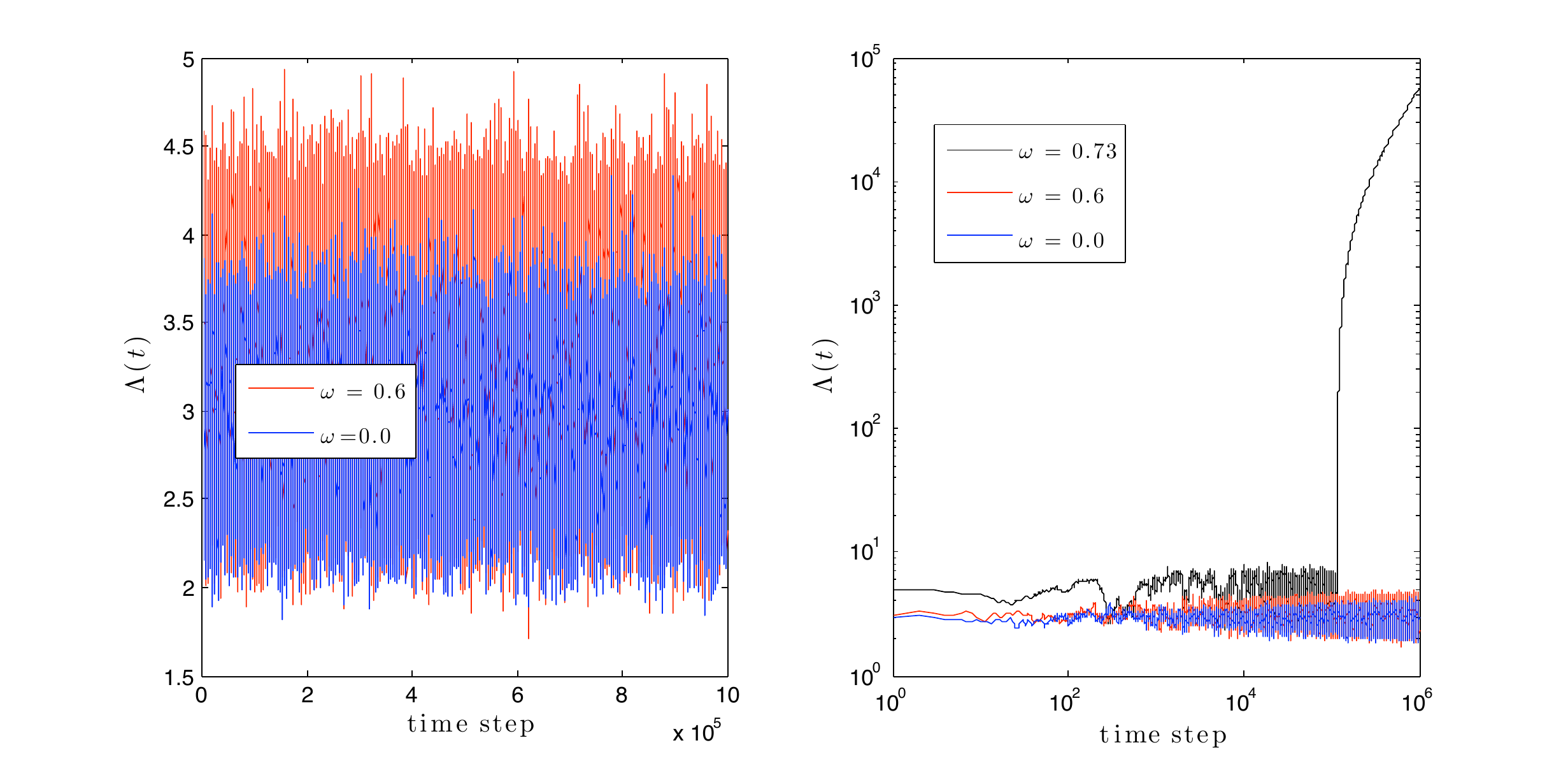}
 \caption{Na\"{i}ve vs. Informed elongation dynamics for $N=10$ and  $1\times10^6$ time steps. In both panels the blue graph corresponds to an elongation time series (\ref{eq:groupElongation}) from a configuration with no informed individuals (called a na\"{i}ve group). In both panels, the red graph shows the result of introducing a single informed individual, where $\omega=0.6$. In the right panel, the black graph also corresponds to a configuration with a single informed particle, but for a higher value of the coupling constant $\omega = 0.73$.  In both panels the group elongation $\Lambda (t)$ remains bounded for all observed times for the naive configuration and the mild coupling ($\omega=0.3$), indicating a configuration that moves cohesively. However,  further increasing the coupling strength (black graph, $\omega=0.73$) causes the group to split, as evidenced by an elongation that grows without bound.}
 \label{fig:lambdat}
\end{center}
\end{figure}
 \begin{eqnarray}
   \label{eq:centroidDef}
   \xb (t)=\frac{1}{N}\sum_{k=1}^N \x_k(t) ,
 \end{eqnarray}   
 and the third one is the mean group velocity  $\vba(t)$ or group polarization                                    
   \begin{equation}
   \label{eq:centroidvDef}
   \vba (t)=\frac{1}{N}\sum_{k=1}^N z_k(t).
 \end{equation}
  \\
In our simulations we observed that there is a non-trivial region of  parameter space that preserves cohesive collective motion (\ref{eq:cohesiveMotion}) for both na\"{i}ve (no informed individuals present, $N_{\rb}=0$) and informed (at least one informed individual present, $N_{\rb}>1$) configurations.  
Figure \ref{fig:lambdat} shows two scenarios for the dynamics of the elongation $\Lambda(t)$.  The left panel shows the situation where the cohesive collective motion property is preserved for two swarms of the same total population size ($N=10$). Blue shows the elongation associated with a na\"{i}ve configuration, and red shows a configuration that includes an informed particle ($\omega = 0.6, N_\beta =1$), observed during $1\times 10^6$ time steps.  We see that the elongation associated with the configuration involving an informed individual tends to take higher values than in the na\"{i}ve case, but remains bounded.  The right panel shows the effect of further increasing the coupling constant,  ($\omega=0.73$, black line) where the elongation remains bounded for some time (about $1\times 10^5$ time steps) after which it starts to increase, signaling that the group has broken apart. \\
\begin{figure}
\begin{center}
 \includegraphics[width=6in]{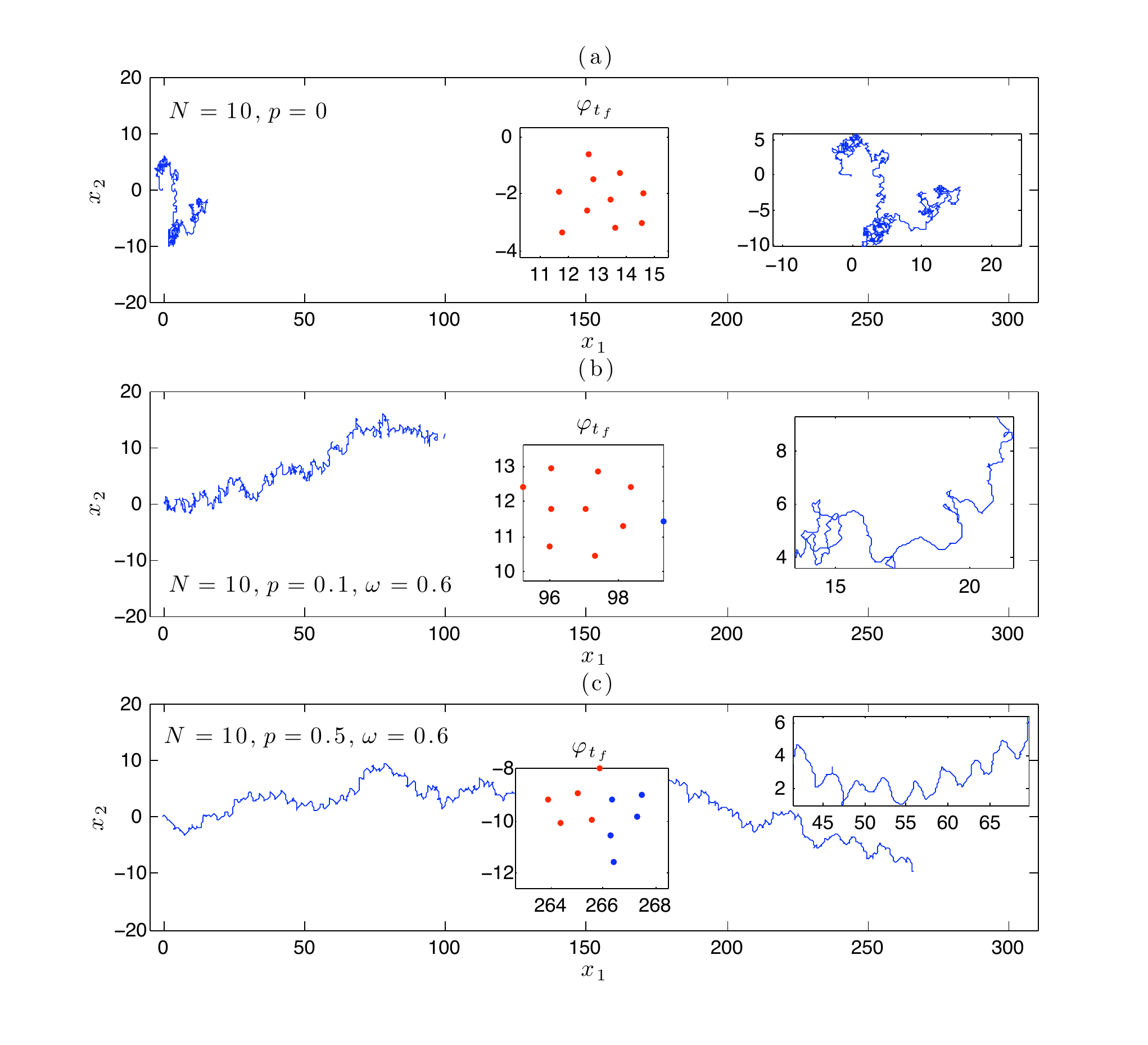}
 \caption{Group centroid sample paths. The blue rugged line in the three panels corresponds to an individual sample path of the configuration centroid $\xb(t)$ up to $t_{f}=1\times 10^{4} $ time steps starting near the origin.  The parameters shared in all three cases are the population size $N=10$, as well as the collective motion parameters, given by  $\sigma=0.1\,\mbox{radians},r_{at}=5.0\,\mbox{m.},\, r_{av}=1.0\,\mbox{m.},\,s=1.0\,\mbox{m./sec.},\, \Delta t = 0.1\,\mbox{secs.}$ In (a) all the individuals are na\"{i}ve, in (b) there is one informed individual with coupling constant $\omega = 0.6$, and in (c) there are 5 informed individuals, also with $\omega=0.6$. The preferred direction is the positive $x_1$ axis.}
 \label{fig:swarm_config}
 \end{center}
\end{figure}
Figure \ref{fig:swarm_config} shows typical sample paths (blue lines) for the configuration centroid  $\xb(t)$ for a group of ten individuals and $T=1\times 10^{4}$ time steps.  The full configuration at the end of the simulation is shown in the insets at the center of each panel, where red dots represent the locations of na\"{i}ve individuals, and blue the informed ones.  Panel (a) corresponds to a configuration involving only na\"{i}ve individuals ($N_{\rb}=0$), Panel (b) has one informed individual ($N_{\rb}=1$) with a coupling constant $\omega=0.6$.  Finally, the lower panel (c) shows the case where five informed individuals are present ($N_{\rb}=5,\omega=0.6$).  The insets to the right show the same sample path over smaller spatial scales.  In the inset of panel (a) we se evidence of  separate clusters --where the group moves very slowly due to a lack of polarization (the slip phase)-- connected by advective flights due to bursts of phase alignment (the stick phase).  This behavior signals the slip/stick dynamics characteristic of these systems \cite{kolpas07}, and resembles the behavior of tracer transport in porous media with a preferential flow direction \cite{berkowitz06}, where the corridors that confine the tracer play a roughly similar role to the polarization bursts in SPP models that lead to ballistic flights, alternating with traps that slow the tracer motion, akin to the slip phase in the SPP. Panel (b) shows the result of adding one informed individual (blue dot) with a relatively high value of the coupling constant ($\omega=0.6$) where without loss of generality we identified the preferred direction with the positive $x_{1}$ axis. The introduction of a single individual is enough to break the orientation symmetry of the na\"{i}ve  case,  resulting in a motion bias along the preferred direction, and an disentanglement of the clusters that appear in the na\"{i}ve case.  Adding more individuals for the same coupling constant leads to a higher mean velocity. In addition to this, the motion develops an oscillatory behavior along the coordinate perpendicular to the informed direction.\\
\begin{figure}
 \begin{center}
 \includegraphics[width=6in]{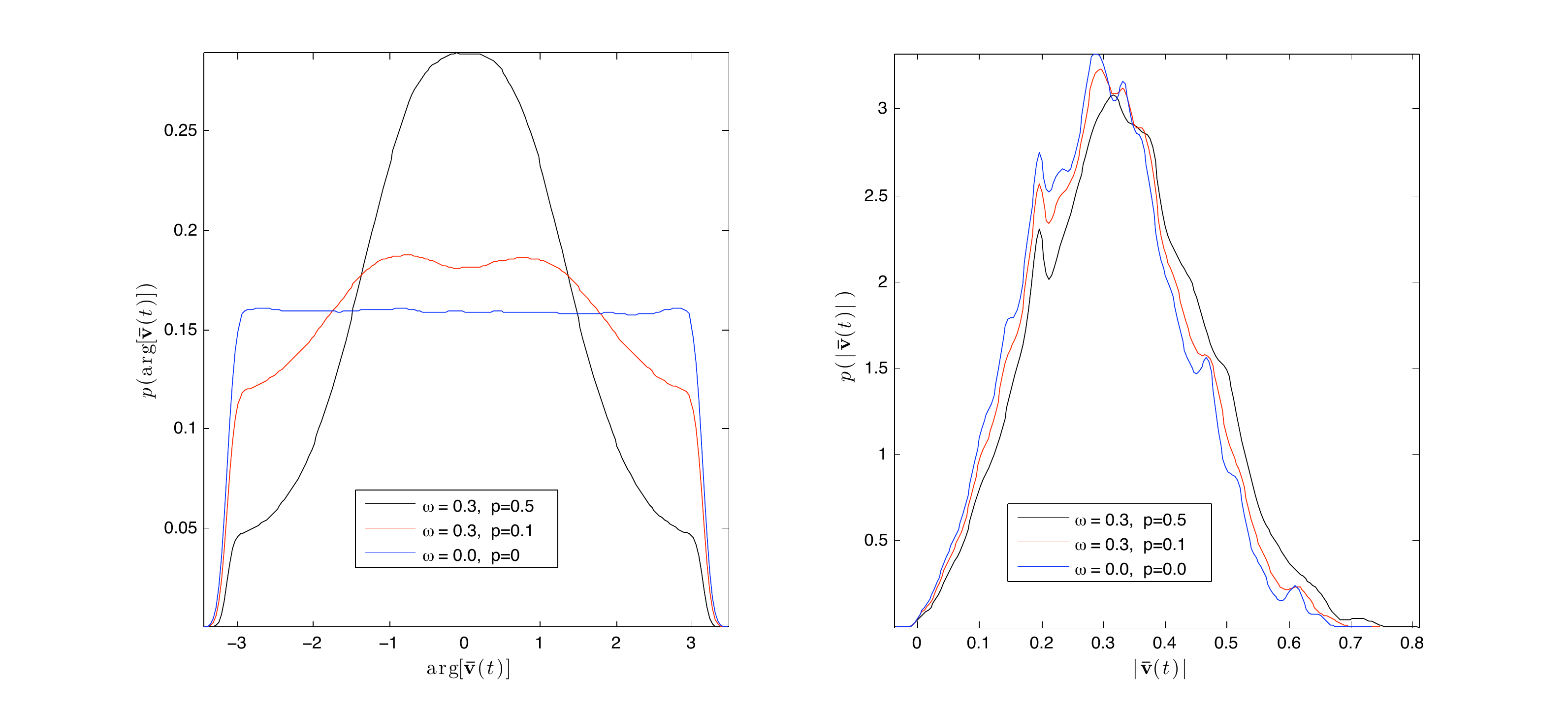}
 \caption{Empirical argument (left) and modulus (right) pdf's for the mean group velocity (\ref{eq:centroidvDef}) for  a configuration of ten individuals.  Both panels show kernel density estimates from a single velocity time series of $3 \times 10^{6}$ data points.  The blue graph corresponds to the na\"{i}ve configuration ($N_{\rb}=0$),  red to a group with one informed individual ($N_{\rb}=1,\omega=0.3$), and black shows the results for a configuration involving five informed individuals, and the same value of the coupling constant ($N_{\rb}=5,\omega=0.3$).}
 \label{fig:pargv_modvt}
 \end{center}
\end{figure}
Figure \ref{fig:pargv_modvt} shows kernel density estimates of the probability density of realized mean group velocities $\vba$ obtained from a single time series ($T=3\times10^6$ time steps) collected after a transient of $1\times10^3$ time steps, for swarms of ten individuals and three different informed regimes. The left panel corresponds to the probability density of mean group orientations $\arg(\vba)$, and the right panel to the modulus (or mean group speeds) $|\vba|$. We see that in the na\"{i}ve swarm (blue graph in the left panel), mean group orientations are chosen uniformly from $[-\pi, \pi)$ at all times.  However, this rotational symmetry is broken upon the introduction of a single informed individual, which yields a symmetric density centered around the informed direction (red, left panel). The existence of peaks reflects a tension between the slip/stick dynamics and the biased motion along the informed direction $\rb$. Overall, the group moves along the informed direction, but slip/stick bursts are strong enough to partially counter that bias by trying to recover the rotational symmetry.  Further increasing the number of individuals (black, left panel) leads to a unimodal density concentrated around the informed direction.   Kernel density estimates for the mean group speeds shown in the right panel $\vba$ show comparatively less variability between the na\"{i}ve and informed regimes. This is to be expected, since the mean group speeds arises mainly from the social rules, and the steering rule is designed to introduce an orientation bias, but has little effect on the modulus.  There is however a tendency to move with higher speeds  as informed individuals are introduced. Critically, the range of variability in speeds is very wide and practically covers the full range of possible values (the individual particle speed is $s=1$ m/s, which constitutes an upper bound for the mean group speed).  \\

\subsubsection{Coarse variables and collective behaviors}
\begin{figure}
 \begin{center}
 \includegraphics[width=6in]{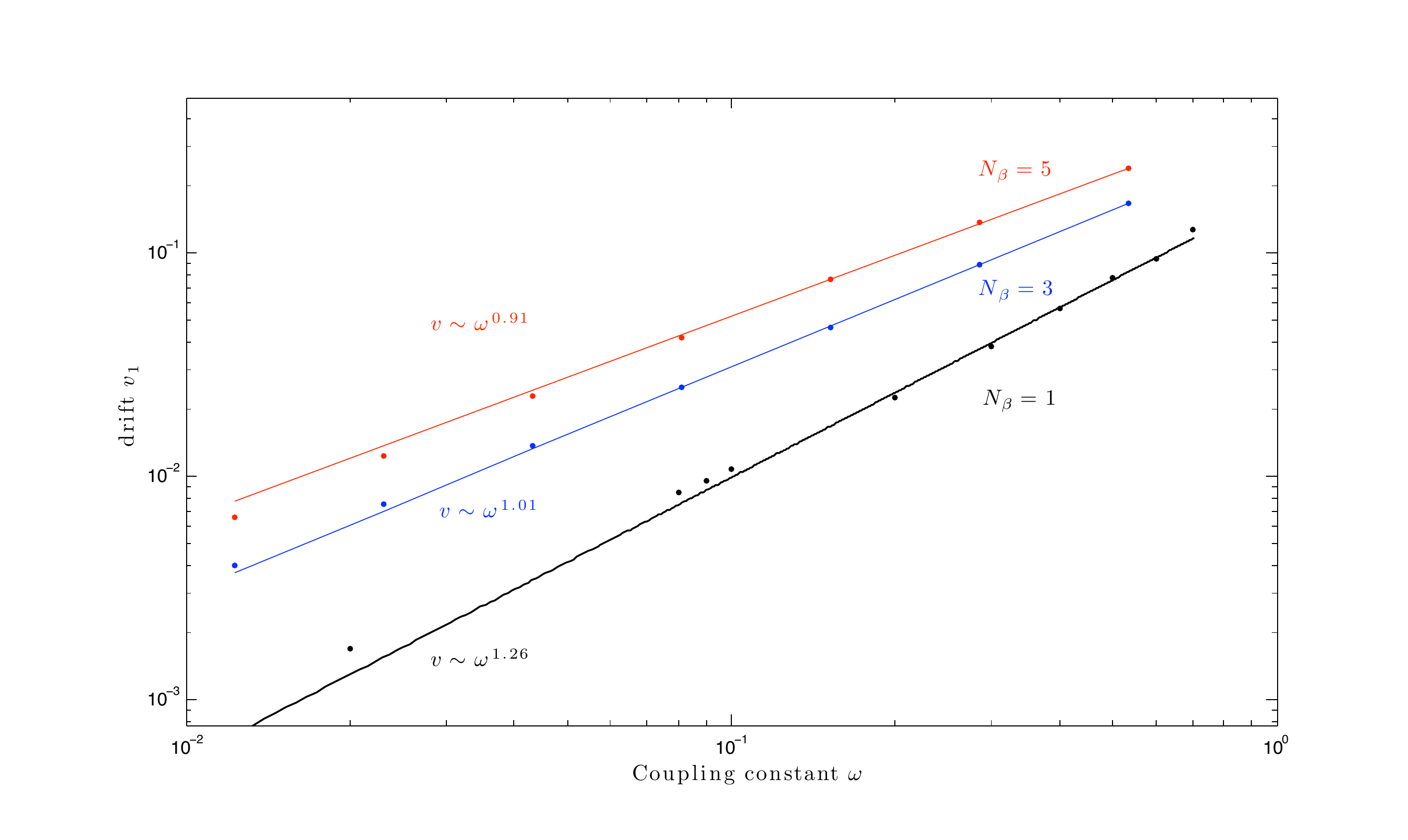}
 \caption{Dependence of the drift velocity $v_1$ along the informed direction $x_{1}$ on the coupling constant $\omega$ for various values of the number of informed individuals $N_\rb=1$ (black), 3 (blue) and 5 (red) for a swarm of $N=10$ individuals.  In all cases, the drift increases as a power law of the coupling constant, but the exponent decreases as the number of informed individuals increases,  due to the upper bound of the group velocity imposed by the individual particle speed.}
 \label{fig:v_vs_omega}
 \end{center}
\end{figure}

\begin{figure}
\begin{center}
 \includegraphics[width=6in]{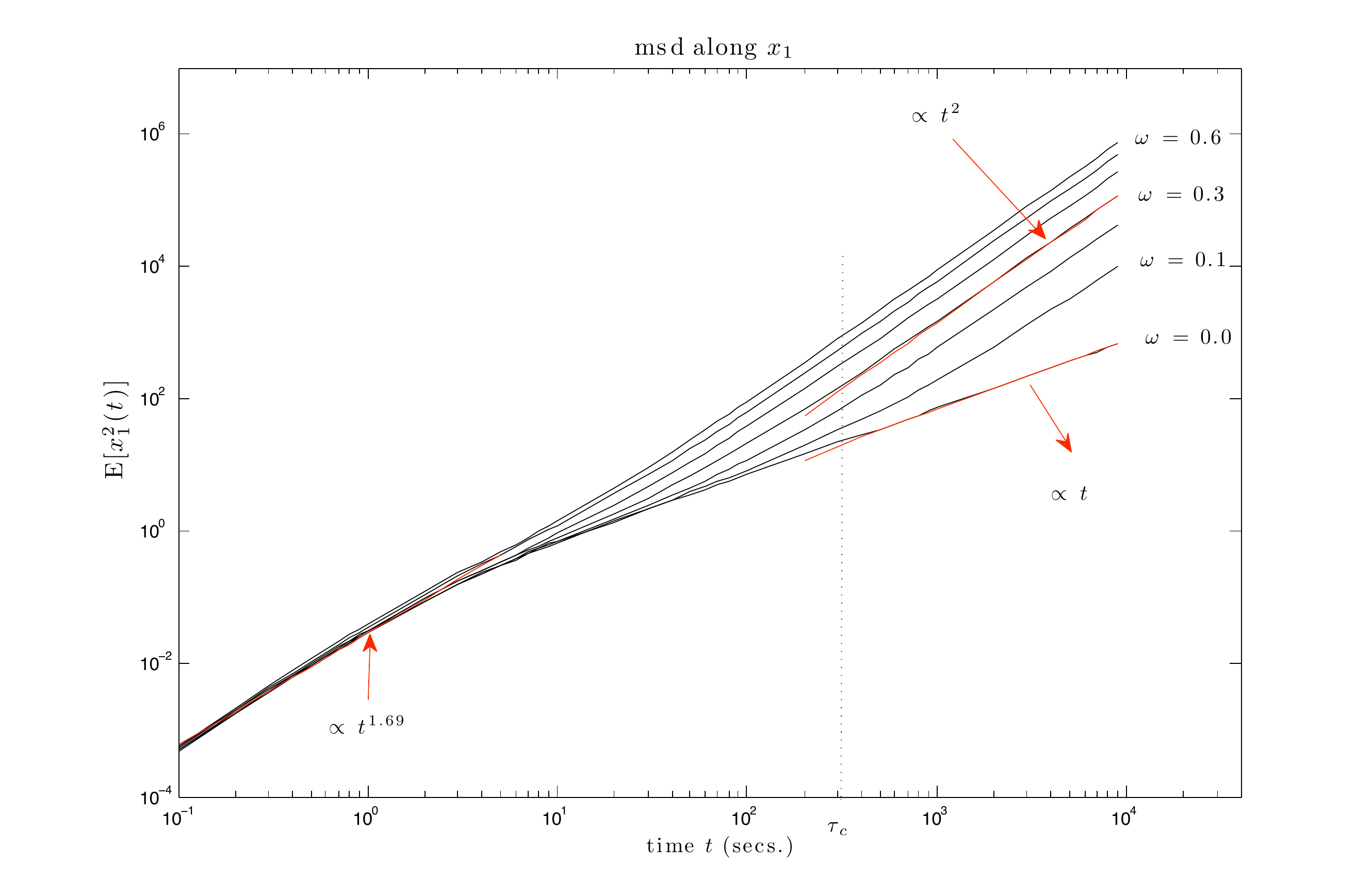}
 \caption{Mean squared displacement along the informed direction $x_1$ versus time for various values of the coupling constant $\omega$, for a total group size $N=10$ and one informed individual for non-zero values of $\omega$.  There is an anomalous transport regime at early times evidenced by the scaling exponent $\alpha \approx 1.7$ (\ref{eq:momrb_mu}).  This regime decays asymptotically to classical diffusion (linear scaling) in the na\"{i}ve configuration.   When $\omega$ becomes positive the transport becomes advective at a characteristic time scale $\tau_c$, signaled by the quadratic scaling of the msd, this time scale becomes shorter with increasing coupling strength. }
 \label{fig:msd_x1}
\end{center}
\end{figure}

\begin{figure}
\begin{center}
 \includegraphics[width=6in]{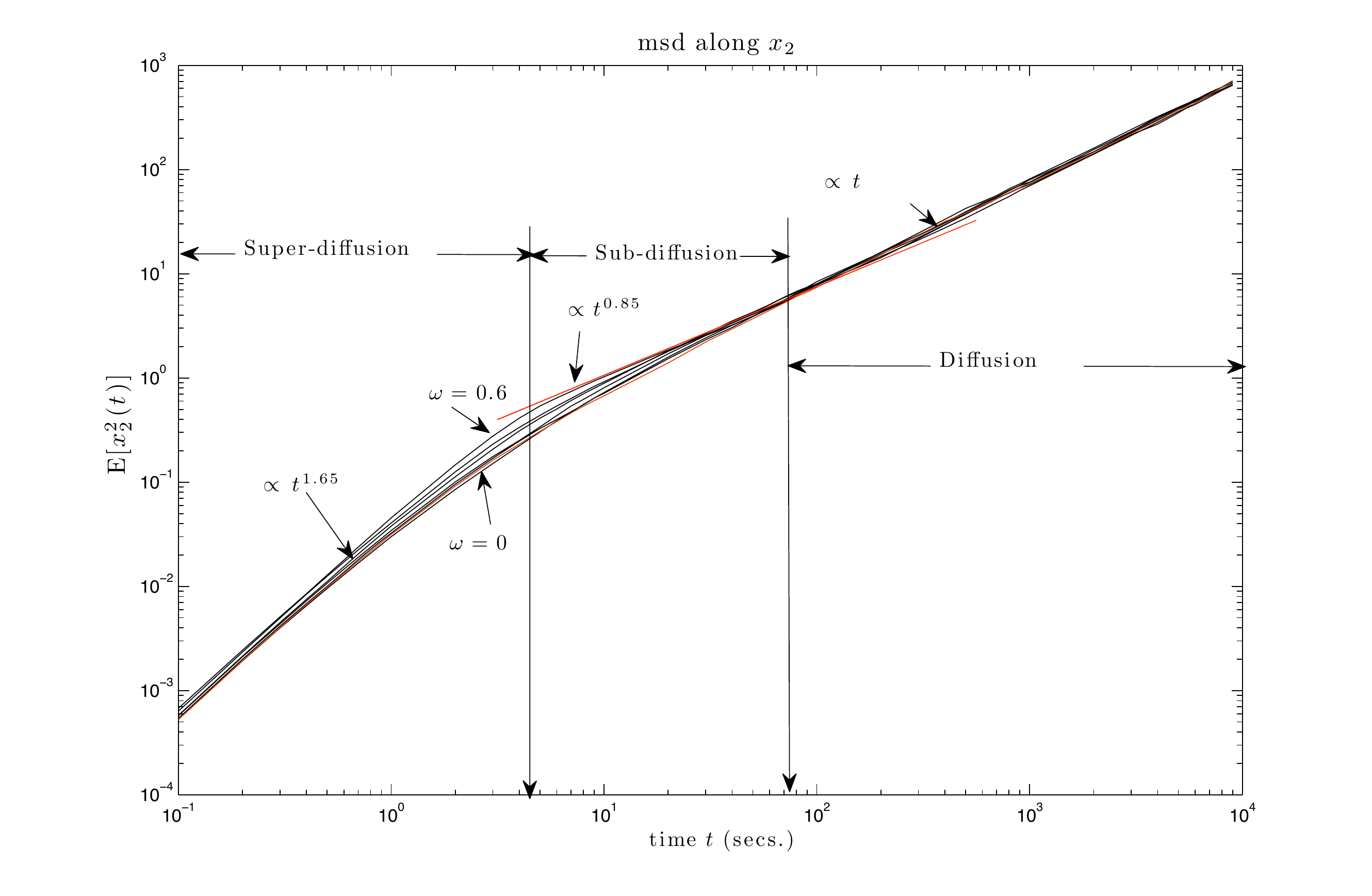}
 \caption{Mean squared displacement along the $x_2$ coordinate versus time.  There is a superdiffusive transport regime at early times  followed by a \emph{sub-diffusive} phase that appears at relatively high values of the coupling constant $(\omega>0.3$).   The anomalous transient gives way to  diffusive transport asymptotically in all cases with numerically indistinguishable diffusion coefficients. }
 \label{fig:msd_x2}
 \end{center}
\end{figure}

\begin{figure}
\begin{center}
 \includegraphics[width=6in]{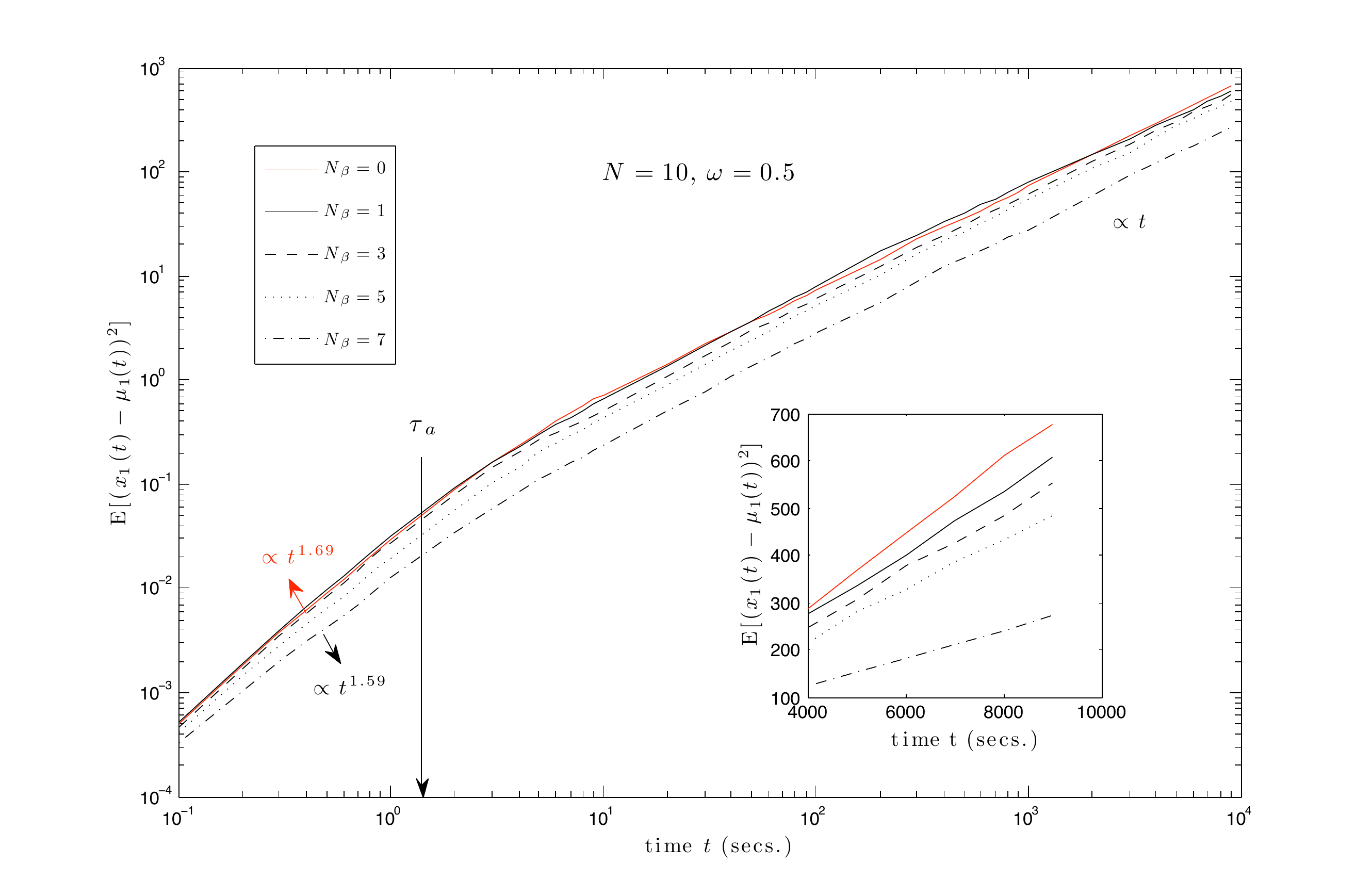}
 \caption{Mean squared displacement with drift removal (\ref{eq:momrb_mu}) along the informed direction $x_1$ versus time for various values of the informed sub-population size and a fixed value ($\omega=0.5$) of the coupling constant and total population size ($N=10$).  The super-diffusive transient eventually gives way to classical diffusion in all cases at a characteristic time scale $\tau_a$.  Both the scaling exponent $\alpha$ and the asymptotic diffusivity decrease as the number of informed individuals increases, which signals an increase in the precision of the decision making  for fixed values of the coupling constant, following the introduction of additional informed individuals while the total group size remains unchanged.}
 \label{fig:msd_mu1}
\end{center}
\end{figure}

Measures for the collective behaviors that are macroscopically relevant can be defined in terms of the scaling properties of the $k$-th moments of the transition probability density $p(\x,t|\mathbf{0},0)$ for finding a configuration meta--particle centroid around position $\x$ at time $t$ given that it started at the origin at time zero. These are defined componentwise by
\begin{eqnarray}
  \label{eq:momrb}
 m_{1}^{(k)}(t) =\E \left[ \bar{x}_{1}^{\,\,k} (t) \right] \sim t^{\delta}, ~~~k \in {1,2}
 \end{eqnarray}
for the $x_{1}$ coordinate, where the scaling exponent $\delta$ determines the prevailing type of transport at the time scale under consideration.
The $k$-th moments (\ref{eq:momrb}) can be calculated from simulations of the individual-based model with the estimator \cite{halmos46}
\begin{equation}
 \label{eq:msdestim}
\hat{m}_1^{(k)}(t)= \frac{1}{Z}\sum_{i=1}^Z \left(X_1^{(i)}(t) \right)^k,
\end{equation}
where $i=1,\ldots, Z$ is the number of simulation runs in the ensemble and each of the $X_{1}(t)=X_{1}(0),X_{1}(1),\ldots,X_{1}(T)$ corresponds to a single time series of length $T$ of group centroid positions along the $x_{1}$ coordinate observed at discrete time intervals of length $\Delta t$. \\
The simplest possible scenario for collective--decision making occurs when the transfer of the bias of the informed sub--population leads on average to motion with effectively constant velocity $v_{1}$ along the informed direction, which we identify without loss of generality with the positive $x_{1}$ axis.  In this case the mean displacement ($k=1$, in (\ref{eq:momrb})) scales linearly with time after a transient determined by the characteristic time scale $\tau_c$, the \emph{time to consensus} 
\begin{equation}
\label{eq:m1_drift}
 m_{1}^{(1)}(t)\sim v_1\,t,
 \end{equation}
\begin{figure}
 \begin{center}
 \includegraphics[width=6in]{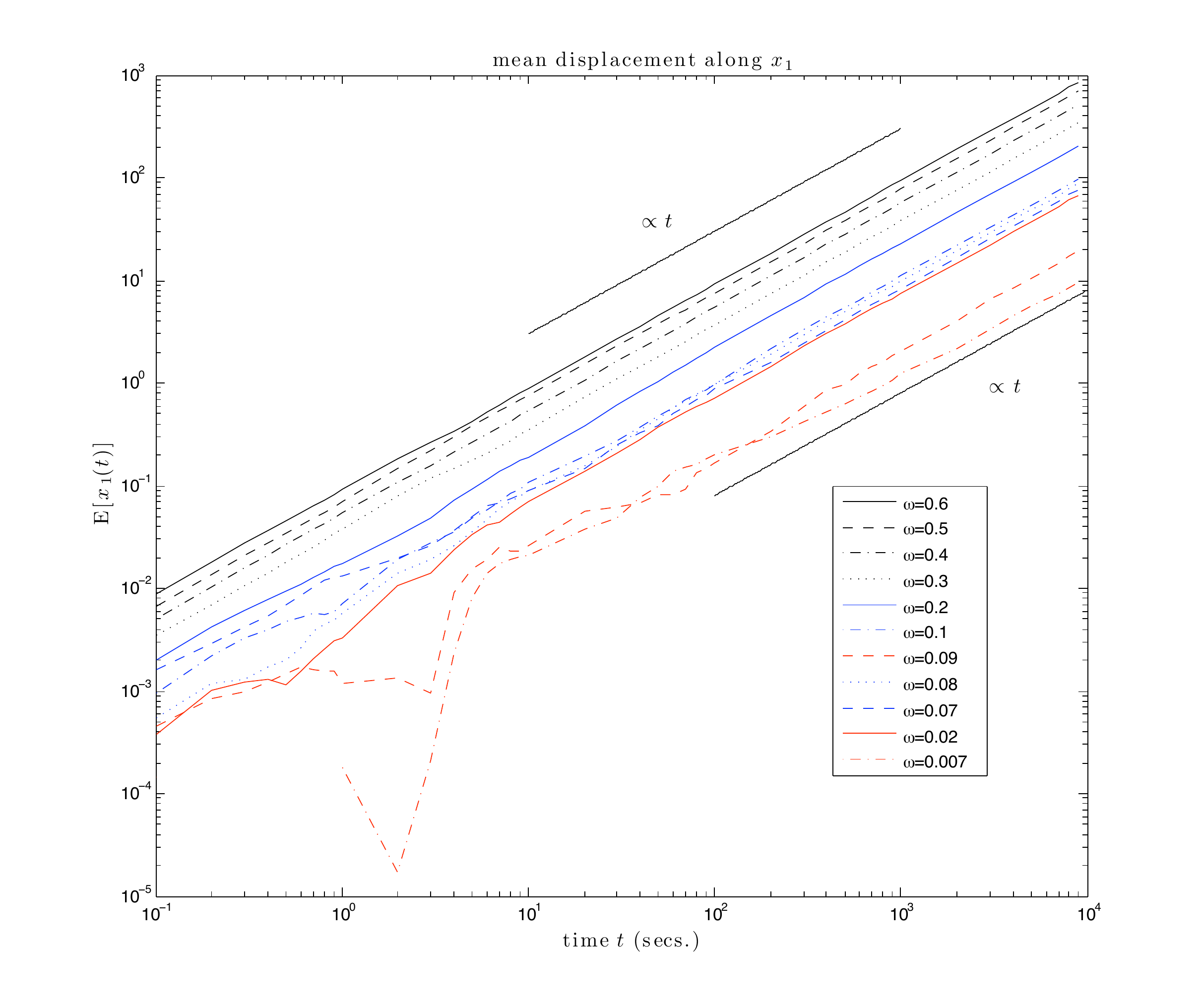}
 \caption{Mean displacement $m_1^{(1)}(t)$ along the informed direction $x_1$ versus time for a wide range of values of the coupling constant $\omega$.   Results are averaged over $3\times10^3$ independent simulation runs.  Initial configurations are given by  uniformly distributed locations within a circle of radius 0.5, and  uniformly distributed orientations on the unit circle. In all cases, the mean displacement increases asymptotically linearly with time, indicating motion with a constant effective speed. The dip at early times for $\omega=0.007$ (red dash-dot line, first from the bottom upwards) constitutes a signature of the transient; however a larger ensemble is required in this case is required to capture it accurately, since the values involved are much smaller than those present for higher values of $\omega$. However, for our purposes, the linear long time behavior is clearly shown. The simulation parameters  are $N=10$ individuals, $N_\rb=1$,$\sigma=0.1$ radians, $r_{at}=5.0$ m, $r_{av}=1.0$ m, $s=1.0$ m/sec.}
 \label{fig:mx1}
 \end{center}
\end{figure}
where the degree of consensus $c$ can be defined by the ratio of the mean group velocity $v_1$ to the individual particle speed $s$
\begin{equation}
\label{eq:consensus}
c = \frac{v_1}{s}.
\end{equation} 
Values of $c$ close to one result from a distribution of individual particle orientations concentrated around the informed orientation.  On the other hand,  if these tend to be distributed uniformly on $(0,\pi]$, one should expect comparatively smaller values of $c$, since the individual particle velocities tend to cancel each other in this regime, which we associate to poor consensus. \\
Figure \ref{fig:v_vs_omega} shows the behavior of the effective velocity $v_1$ versus the coupling constant $\omega$ for a SPP swarm of ten particles and various proportions of informed individuals. We see that in all cases  the degree of consensus increases as a power law of the coupling constant $\omega$  with a slope that decreases as informed individuals are added.   This means that if there is an optimum group velocity in some appropriately defined sense,  it can be reached collectively by two different avenues.  One is to have a small number of informed individuals at a high coupling constant, and the other is to have a large number of informed particles with low values of the coupling constant.  The power law dependence implies that the difference in values of the coupling constant for these two strategies can span orders of magnitude.   Therefore, if the cost of recruiting informed individuals is less than that of leading the group, a possible optimal strategy, would consist of recruiting additional informed individuals, each of them with comparatively smaller values of $\omega$, instead of simply increasing the coupling constant of the informed group size, which comes at the additional complication of increasing the probability of having the group split apart. \\
Naturally, these two strategies to reach the same target group velocity, are likely to have different accuracies.  This can be more readily detected in measures of spread along the mean value, like the second moment. For this purpose we use the msd ($k=2$, in (\ref{eq:momrb})), which  can also detect very efficiently the various types of collective behaviors, either transient or asymptotic,  that contribute to macroscopic transport.  For instance, the time to consensus can be detected sharply by a transition from linear (diffusion-dominated) or anomalous scaling to a \emph{quadratic} one at the point in time where advection begins to dominate
\begin{equation}
\label{eq:m2_drift}
 m_{1}^{(2)}(t)\sim v_1^2\,t^2.
 \end{equation}
Figure \ref{fig:msd_x1} shows the msd along the informed direction $x_{1}$ for a group of 10 individuals, one of them informed, and various values of the coupling constant.  There is a transition from linear to quadratic scaling for non--negative values of $\omega$ at a characteristic time scale $\tau_c$.  As one increases the coupling constant, the informed sub--population becomes more efficient at transferring their bias to the whole group, signaled by an earlier time to consensus.    The transient regime appears to be anomalous (supperdiffusive) in both na\"{i}ve ($\omega=0$) and informed ($\omega>0$) configurations.  The anomalous transient can be better detected by looking at the scaling of the second order fluctuations, which requires removing the mean value in the definition of the moments (\ref{eq:momrb}) for $k=2$,
\begin{equation}
\label{eq:momrb_mu}
 M_1^{(2)}(t) =\E \left[ \left( \,\bar{x}_1(t) - m_1^{(1)}(t) \, \right)^2 \right] \sim t^{\alpha}, 
 \end{equation}
If  $\alpha=1$ in (\ref{eq:momrb_mu}) for some set of time scales, then the fluctuations behave as classical diffusion \cite{gar85,van01}.   Values of $\alpha$ different from one are dubbed \emph{anomalous} and can be of two main types: sub--diffusive or `trapped-diffusion' for $0<\alpha<1$, and super-diffusive (sub--ballistic)  or `enhanced diffusion',  if $1<\alpha < 2$.  These anomalous behaviors signal the presence of fluctuations that have persistent  correlations in space, time, or both at macroscopically relevant scales \cite{west03,metzler04}.  Slip/stick dynamics dominate the early time behavior of the second moment of the fluctuations along the $x_1$ (Figure \ref{fig:msd_mu1} ) and $x_2$  (Figure \ref{fig:msd_x2}) coordinates, where the transport is clearly anomalous. There is a (sub-ballistic) super-diffusive transient that eventually decays to classical diffusion at a characteristic time scale $\tau_a$.  This is due to the alternation of bursts of advective flights (the slip phase) due to polarization of the orientations, that are interrupted when the polarization is lost and the group moves much more slowly (stick).  Since the msd eventually becomes diffusive, the  temporal correlations in the mean velocity induced by the polarization eventually decay at a characteristic time scale $\tau_{a}$ (shown in Figure \ref{fig:msd_mu1}), after which the fluctuations are diffusive, with identical diffusion coefficients along both coordinates. \\
The role of the informed sub-population is more nuanced along the $x_2$ coordinate  (see Figure \ref{fig:msd_x2}), in the sense that for higher values of the coupling constant, there is a clearly detectable  \emph{sub-diffusive} regime between the early time super-diffusive transient and the diffusive regime; this is due to reversals in velocity that are more marked along the $x_{2}$ direction.  Thus, the introduction of informed individuals at high coupling constants induces an anisotropy not only in the values of the diffusion coefficients but also in the manner in which the correlations decay.  Whereas the the ADEM (\ref{eq:ADEM1}) allows for anisotropies in the diffusion coefficients along the two directions, the memory kernel $M(t)$ must be identical along each direction.  This precludes the use of the ADEM (\ref{eq:ADEM1}) for predicting the behavior of the full 2--dimensional density, which would require a separate treatment along each coordinate.  We can still however use it for the marginals along each  direction in order to predict the mean squared displacements, and to extract the transport coefficients.\\

Other macroscopic measures of interest are related to the precision of the decision-making process.  This quantity is directly linked to the strength of the fluctuations along the informed direction, relative to those of a fully na\"{i}ve swarm of the same total population size.  Higher precision in decision-making naturally implies a distribution of centroid positions along $x_1$ that is highly concentrated around the mean value.  This is measured by the magnitude of the diffusivity along  $x_1$.  Figure \ref{fig:msd_mu1} shows estimates of the second order fluctuations along the informed direction for various values of the number of informed individuals $N_\rb= 0, 1, 3, 5, 7$ and a fixed coupling strength of $\omega=0.5$.  All the fluctuations scale asymptotically linearly and thus they are dominated by classical diffusion, but  with a diffusivity that decreases as the number of informed individuals $N_\rb$ increases. This agrees with the intuition that the precision of the decision--making process should improve as informed individuals are added to a group of fixed size.  Figure \ref{fig:msd_x2} shows the mean squared displacement along the direction perpendicular to the informed orientation $x_2$ for the same parameters in Figure \ref{fig:msd_x1}. We see that the diffusivity along this coordinate is numerically indistinguishable between na\"{i}ve and informed configurations for the various values of the coupling constant used.  This suggests a definition of decision--making precision given by the ratio of the diffusivities along the two coordinates,
\begin{equation}
\label{eq:precision}
\rho= \frac{D_1}{D_2} \leq 1.
\end{equation}
Values of $\rho$ close to one  indicate a spread of the meta-particle positions that is comparable along both coordinates, and thus poor precision in decision--making, and the opposite situation occurs for values of $\rho$ that are significantly less than one. \\
\section{Continous time random walks and the advection-diffusion equation  with memory} \label{sec:ctrw}
In this section we  briefly review known results about continuous time random walk models (CTRW), that were originally introduced to describe the random motion of a particle on a disordered lattice (or medium).  The main innovation of CTRW theory consisted in allowing  the lattice spacing and updating times to become random variables themselves.  It can be shown \cite{montroll73,kenkre77} that when the distribution of lattice spaces has finite moments of all orders, the evolution of the transition density for the location probability density of a CTRW is given by the generalized master equation (GME)  \cite{kenkre77,west97,kenkre03,kenkre09} also called  the advection-diffusion equation with memory (ADEM) when there is a drift \cite{cortis04,berkowitz06}. This generalization of the classical random walk leads to an evolution equation for the transition probability density of the particle position that is non-local in time, since the flux depends on a weighted time average over the full past.  The weighting function is commonly called  a `memory function', and results from the wide range of transition rates originating from the spatial disorder.  This approach has been successfully applied in models of anomalous diffusion \cite{metzler04}, that typically require the memory to decay algebraically instead of exponentially.   Approaches based on the CTRW and GME methods however are more general, and allow for any functional form of the memory, provided that it can be normalized.  \\
As will be discussed in Section \ref{sec:multiscale}, the memory function plays  a fundamental role, since it encodes macroscopic transport coefficients of interest, together with their characteristic time scales. We will exploit these properties of memory functions in order to estimate the  transport parameters associated with the various types of collective behaviors arising in SPP models of collective motion, with and without informed individuals. 
\subsection{Continuous time random walks}\label{ssec:ctrw&adem}
Continuous time random walks (CTRW) \cite{montroll73,kenkre77,klafter80}  are a generalization of classical random walks \cite{gar85,van01}  where  the jump size $\Delta x$ and updating time $\Delta t$ are allowed to become random variables.  Sample paths are generated by drawing the jump size $\xi$, and waiting time $\tau$  from the joint probability density $\psi(\xi,\tau)$. The elapsed time $t_n$ for such a walker after $n$ steps is,
\[
t = \sum_{j=1}^{n}\tau_j,~~~\tau_j \in \mathbb{R}^+,
\]
and the position $\x_n(t)$, for a 2-dimensional walk,
\[
\x_n(t)=\sum_{j=1}^n  \xi_j,~~~ \xi_j \in \mathbb{R}^2.
\]
The probability of observing a walker at position $\x$ at time $t$ given that  it started at the origin at time zero is,
\begin{equation}
\label{eq:chapmanCTRW}
p(\x,t) = \delta(\x)\,\Psi(t) + \int_{\mathbb{R}^2}\int_0^t \psi(\xi,\tau)\,p(\x-\xi,t-\tau)\,d\xi\,d\tau,
\end{equation}
where the survival function $\Psi(t)$ is the cumulative of the waiting time marginal density of $\psi(\xi,\tau)$ 
\begin{equation}
  \label{eq:survival}
   \Psi(t) = 1-\int_{\mathbb{R}^2}\int_0^t \psi(\xi,\tau)\,d\xi \, d\tau.
\end{equation}
For the particular situation were the jumps and waiting times are decoupled, the joint density $\psi(\xi,\tau)$ can be rewritten as $\psi(\tau)\lambda(\xi)$, where $\psi(\tau)$ is the distribution of waiting times, and $\lambda(\xi)$ is the distribution of jumps.  This assumption simplifies (\ref{eq:chapmanCTRW}) to
\begin{equation}
\label{eq:uncoupledchapman}
p(\x,t) = \delta(\x)\,\Psi(t) + \int_{\mathbb{R}^2}\lambda(\xi)\int_0^t \psi(\tau)\,p(\x-\xi,t-\tau)\,d\xi\,d\tau.
\end{equation}
If the jump density $\lambda(\xi)$ has finite moments, and $p(\x,t)$ can be expanded in a Taylor series, it can be shown \cite{berkowitz06} that the differential version of (\ref{eq:uncoupledchapman}) corresponds to an advection--diffusion equation generalized to non-local time, 
\begin{eqnarray}
 \label{eq:ADEM}
 \frac{\partial \pa}{\partial t}&=&-\int_{0}^{t}\,M(t-s)\,\left[ \vb \cdot \nabla p(\x,s) - \Db\,  
: \,  \nabla \, \nabla p(\x,s)\right] \, ds \\  
\nonumber p(\x,0^{+}) &=& \delta(\x), ~~\x \in \re^{2},~~t\in \re^{+}
 \end{eqnarray}
where the memory term $M(t)$ can alternatively be defined in terms of the Laplace transform of the waiting time density, or  as the kernel of the velocity time autocorrelation (divided by 2D so that it integrates to one).  In the former case we have,
\begin{equation}
\widetilde{M}(\epsilon)=\frac{\bar{t} \epsilon \tilde{\psi}(\epsilon)}{1-\tilde{\psi}(\epsilon)}
\label{eq:mem_lpl}
\end{equation}
 and $\tilde{f}(\epsilon)$ denotes the Laplace transform of a function $f(t)$ with $\epsilon$ being the Laplace variable, and $\bar{t}$ is the characteristic time between transitions
 \begin{equation}
 \label{eq:tbar}
 \bar{t}=\int_0^\infty \tau\,\psi(\tau)\,d\tau.
 \end{equation}
The drift  term $\vb$ in (\ref{eq:ADEM}) is related to the first moment of the jump pdf $\lambda(\xi)$,
\begin{equation}
 \label{eq:drift}
 \vb =\frac{1}{\bar{t}} \int_{\mathbb{R}^{2}}\x\,\lambda (\x) d \x,
\end{equation}
and the diffusivity tensor $\Db$ is given by the second moment of $\lambda(\xi)$
\begin{equation}
 \label{eq:diffusivity}
 \Db =\frac{1}{2 \,\bar{t}} \int_{\mathbb{R}^{2}}\x\, \x^T\,\lambda (\x) d \x.
\end{equation}
In the drift-free case, the Laplace domain solution of the ADEM (\ref{eq:ADEM}) is given by
\begin{equation}
\label{eq:sln_dem_lpl}
\tilde{p} (\x,\epsilon)=\frac{1}{2\,\pi \,\widetilde{M}(\epsilon)\sqrt{D_1\,D_2}}  \,K_0\left( \sqrt{ \frac{
                                   \epsilon
                                   }{
                                   \widetilde{M}(\epsilon) 
                                   } 
                          \left[         
                              \frac{x_1^2}{D_1}+\frac{x_2^2}{D_2}  
                         \right]   
                        } 
                              \,  \right),
\end{equation}
which assumes that the off--diagonal components of the diffusivity tensor are zero,  $D_1$ and $D_2$ are the diffusivities along the $x_1$ and $x_2$ coordinates, $K_0$ is the modified Bessel function and 
$\tilde{M}(\epsilon)$ is the Laplace transform of the memory with $\epsilon$ being the Laplace variable. If the drift vector has a single non-vanishing component which coincides with the $x_1$ direction, the solution is  \cite{berkowitz06}
\begin{equation}
 \label{eq:sln_adem_lpl}
\tilde{p} (\x,\epsilon)=\frac{1}{2\pi\, \widetilde{M}(\epsilon) \sqrt{D_1D_2}} \exp\left(\frac{x_1\,v_1}{2D_1}\right)\, K_0 \left( 
                                                                 \frac{v_1}{2\,D_1}
                                                                 \sqrt{
                                                                 x_1^2+\frac{D_1}{D_2} x_2^2     
                                                                      \left[1+4\frac{\epsilon\, D_1
                                                                                          } {
                                                                                          \widetilde{M}(\epsilon) \,v_1^2
                                                                                          }
                                                                        \right]
                                                                        }
                                                                     \right),
\end{equation}
where $v_1$ is the magnitude of the drift along the $x_1$ coordinate.  Finally, the Laplace domain expression for the mean squared displacement of the ADEM (\ref{eq:ADEM}) along the direction of the drift is
\begin{equation}
\label{eq:msd_lpl}
\tilde{m}^{(2)}_1(\epsilon) = \frac{2\,v_1^2}{\epsilon^3}\, \widetilde{M}_1^2(\epsilon) + \frac{2\,D_1}{\epsilon^2}\, \widetilde{M}_1(\epsilon),
\end{equation}
which yields all the characteristic time scales after Laplace inversion. \\
\section{Multiscale Method}
\label{sec:multiscale} 
We start with the assumption that the meta-particle random walk follows an unknown  CTRW with independent jump and waiting time distributions.  In this case one can assert that after the velocity--autocorrelation equilibrates, the evolution of the transition density for the meta--particle location $p(\x,t|\mathbf{0},0)$ is given by an advection--diffusion equation with memory (\ref{eq:ADEM}) under relatively mild assumptions.  The ADEM would be fully specified if analytical forms of the jump and waiting time densities were known on the basis of the SPP formulation.  Unfortunately, this is not the case.   We instead \emph{estimate} them from a single velocity and centroid position time series obtained from a simulation run of the spp model with a combination of non--parametric and parametric methods.  We show that this simple estimation procedure predicts mean squared displacements that are indistinguishable numerically from those estimated from an ensemble average over a large number of simulation runs.  We exploit these results in order to explore a wide region of the parameter space, and obtain analytical results for the time to consensus based on the functional forms used in the parametric estimation of the memory.  These results are exact for the case of exponential and Gamma density (\ref{eq:gammamem}) memories, but only approximate for the truncated Mittag--Leffler case (\ref{eq:mlfmem}). \\
\subsection{Estimation of $M(t)$}\label{ssec:estim_phi}
Although the memory in (\ref{eq:ADEM}) is defined in terms of the Laplace transform of the distribution of waiting times (\ref{eq:mem_lpl}), a more convenient definition relates it to the time velocity autocorrelation of the random walker \cite{west03,west97,kenkre09}
\begin{equation}
\label{eq:mem_autocorr}
M(t)=\frac{1}{2(D_1+D_2)}\,\E\left[\left(\va(\tau)-\mu \right)\cdot \left(
\va(\tau+t)-\mu\right)\right]
\end{equation}
where  $\mu$ is the expected value of the random velocity $\va(t)$. The definition of the ADEM (\ref{eq:ADEM}) allows different diffusivities along each coordinate, but not  anisotropies in which the correlations decay differently along each component of the velocity. In general though, one should consider the memory separately along each component for an accurate description of the evolution of the transition pdf.  Unfortunately, this turns out to be the case for informed swarms at high coupling constants, as can by seen after observing the differences for high values of the coupling constant in the mean squared displacements along $x_2$  (Figure  
 \ref{fig:msd_x2}) and the drift-corrected msd along $x_1$ (Figure \ref{fig:msd_mu1}).  The former has a distinct sub-difussive regime that is not apparent in the latter.  For the purposes of collective--decision making, it suffices to focus on the behavior of the  msd (\ref{eq:msd_lpl}) along the informed direction, and the macroscopic transport coefficients $D_1,D_2$ and $v_1$. \\

\begin{figure}
 \begin{center}
 \includegraphics[width=6in]{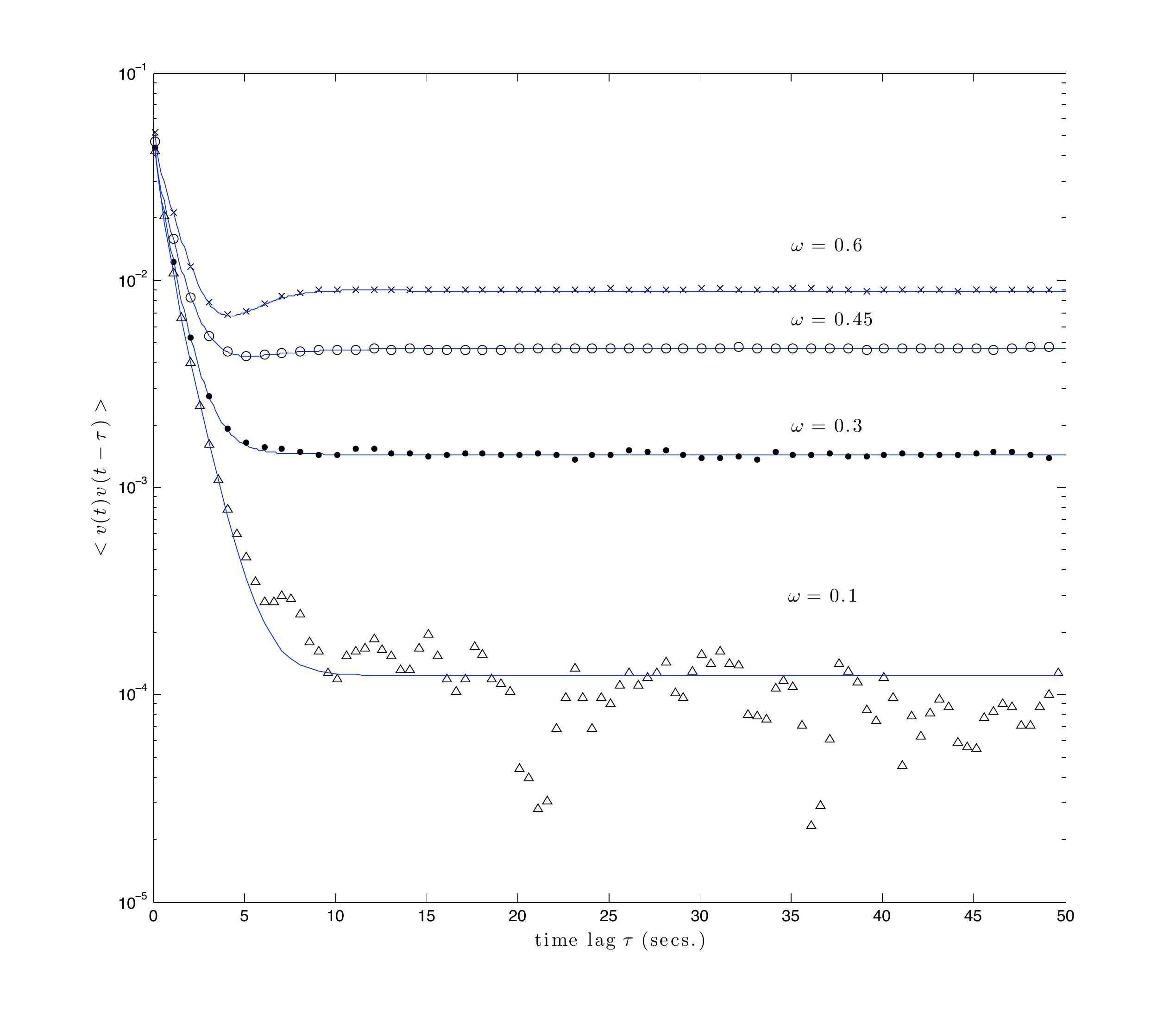}
 \caption{Estimated velocity autocorrelation function from a single time series of $1x10^7$ time steps (marks) and fitted functions (blue lines). The group size in the simulation was 10 individuals, one of them informed.  The black markers correspond to the non-parametric estimates, for various values of $\omega$.
The continuous lines show the parametric fits with a Gamma kernel $f(\tau)+\bar{v}_1^2$ (
\ref{eq:gammamem}) for $\omega = 0.1$ and 0.3, and a truncated Mittag--Leffler function $g(\tau)+\bar{v}_1^2$ (\ref{eq:mlfmem}) for $\omega=0.45$ and 0.6.  Parameter estimates and goodness of fit values can be found in Table \ref{table:parameters} }
\label{fig:mem_simandfits_n10}
 \end{center}
\end{figure}

\begin{figure}
 \begin{center}
 \includegraphics[width=6in]{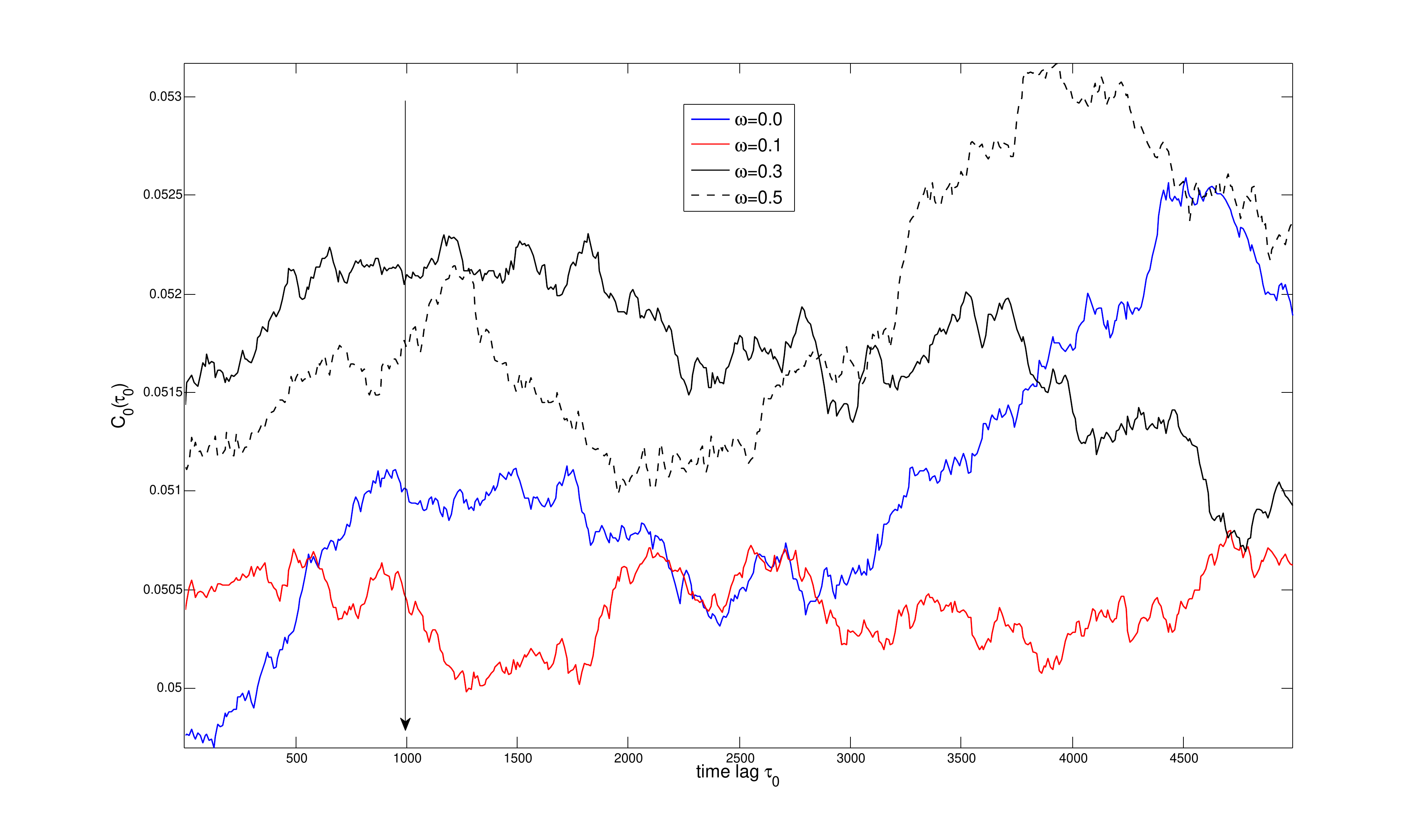}
 \caption{Velocity autocorrelation function at zero lag $C_{0}(\tau_{0})$ computed from a window of fixed length $T=1\times 10^{4}$ time steps, and shifting the origin of the first data point in the window $\tau_{0}$ time steps from the absolute origin of the simulation run. Each graph corresponds to a different value of the coupling constant. The swarm simulation consisted of a total group size of $N=10$ individuals, of which one is informed. The arrow indicates the point at which data started to be collected for the estimates of the macroscopic transport parameters ($\tau_{0}=1000$ time steps.) }
\label{fig:cstationarity}
 \end{center}
\end{figure}

We first compute a non-parametric estimate of the velocity auto-correlation function from a velocity time series  $\{v_1,v_2,\ldots,v_T\}$ obtained from a single simulation run of the SPP, where  each of the $v_i, i=1,\ldots T$ is the component of the meta-particle velocity along the informed direction, sampled at discrete time intervals $\Delta t$, and $T$ is the length of the time series. We used the unbiased estimator \cite{orfanidis96}
\begin{equation}
\label{eq:nonp_c}
\widehat{C}(\tau)=\frac{1}{T-\tau}\sum_{i=1}^{T-\tau}\left(v_i-\bar{v}\right)\,\left(v_{i+\tau}-\bar{v}\right), ~~~ \tau=0,\ldots,T-1,
\end{equation}
where $\tau$ is the time lag and $\bar{v}$ is the sample mean,
\[
\bar{v}=\frac{1}{T}\sum_{i=1}^{T}v_i.
\]
The tabulated function that results from the non-parametric estimate (\ref{eq:nonp_c}) is fed to a non-linear least squares routine that yields a \emph{parametric} estimate of the memory (see below).  The Laplace transform of this function is then substituted into the  expression for the mean squared displacement (\ref{eq:msd_lpl}), or  the transition pdfs (\ref{eq:sln_dem_lpl}) and (\ref{eq:sln_adem_lpl}), all of which can then be inverted numerically.  The parametric estimate requires a `template' function for the velocity auto--correlation that fits the data well and has a known analytical Laplace transform.  We identified two functions that provide remarkably good fit and have very simple transforms. For lower values of the coupling constant this template is the  Gamma density (Figure \ref{fig:mem_simandfits_n10}),
\begin{equation}
\label{eq:gammamem}
f(t) = \frac{\tau_a^{\beta-1}}{\Gamma(1-\beta)}\, t^{-\beta}\,e^{-t/\tau_a}
\end{equation}
where $\tau$ controls the exponential decay, and the exponent $\beta$ controls the initial algebraic decay.  The Laplace transform of (\ref{eq:gammamem}) is simply
\begin{equation}
\label{eq:gamma_lpl}
\tilde{f}_1(\epsilon)=\left(\frac{1}{\tau_{a}}+\epsilon\right)^{\beta-1}.
\end{equation}
The second function is appropriate for higher values of the coupling constant $\omega$ which leads to oscillations (see Figure \ref{fig:mem_simandfits_n10}).  In this function the initial power law decay in the Gamma density in (\ref{eq:gammamem}) is substituted by an exponentially truncated Mittag--Leffler function \cite{podlubny99,west03},
\begin{equation}
\label{eq:mlfmem}
 g(t) =  \frac{\tau_\epsilon+\tau_a^\alpha}{\tau_\epsilon \tau_a^\beta}     \,t^{\beta-1}\,E_{\alpha,\beta}(-t^\alpha/\tau_\epsilon)\,\exp(-t/\tau_a),
 \end{equation}
where the Mittag--Leffler function $E_{\alpha,\beta}(-t^\alpha/\tau_\epsilon)$ is defined as
\begin{equation}
 \label{eq:mlfdef}
 E_{\alpha,\beta}(-t^\alpha/\tau_\epsilon) = \sum_{k=0}^{\infty}\frac{(-1)^k (t^\alpha/\tau_\epsilon)^k}{\Gamma(\alpha k+\beta)}
\end{equation}
where $\alpha$ and $\beta$ are shape parameters and $\tau_\epsilon$ controls the transition between the early time and the asymptotic regime. The Laplace transform of the truncated Mittag--Leffler function
(\ref{eq:mlfmem}) is also very simple \cite{podlubny99}
\begin{equation}
\label{eq:mlf_lpl}
\mathcal{L}\left[\frac{}{}
                         t^{\beta-1}\,E_{\alpha, \beta}(-a\,t^\alpha)\,e^{-b \,t}
                    \right] (\epsilon )=
                     \frac{
                        (b+ \epsilon)^{\alpha-\beta}
                            }{
                                \left(
                               b + \epsilon
                                     \right)^\alpha + a
                               }.
\end{equation} 
Figure \ref{fig:mem_simandfits_n10} shows the results the estimation procedure for a swarm of $N=10$ individuals, one of them informed.  The black marks show the non-parametric estimates based on (\ref{eq:nonp_c}) and the blue lines show the parametric fits using the Gamma density (\ref{eq:gammamem}) for lower values of the coupling constant ($\omega=0.1$ and 0.3), and the truncated Mittag--Leffler function for higher values ($\omega=0.45$ and 0.6).  In all cases, the template functions provide remarkably good fit, including the oscillation that appears for higher values of $\omega$.  The functions eventually decay to a constant value that corresponds to the drift squared $v_{1}^{2}$, which we do not remove from the estimator, in order to be able to resolve the changes in the qualitative behavior of the correlation function for various values of $\omega$.  Table \ref{table:parameters} shows the parameter estimates in all four cases, together with goodness of fit values.
\begin{table}[htdp]
\begin{center}
\begin{tabular}{|c|c|c|c|c|c|c|c|c|}
\hline
$\omega$ & $D_1$     & $v_1$   &    $\tau_\epsilon^{\ast}$  & $\tau_a$  & $\alpha$ & $\beta$ & $R^2$ & SSE  \\ [0.5ex]
\hline
$0.10$   & 0.0373  & 0.012 &   -    &   8.33    & -  & 0.20          & 0.996 & $2.3 \times 10^{-6}$\\ 
\hline
$0.30$   &  0.0373 &  0.038                         &    -    &  0.94   &  -  & 0.24         & 0.998 &$2.9 \times 10^{-7}$\\
\hline
$0.45$ &   0.0326    & 0.068       & 0.55 &   0.26   &  1.09  &  0.86  & 0.999 &$4.4 \times 10^{-7}$\\ 
\hline
$0.60$   &  0.0282      & 0.094       & 0.62&   0.37   & 1.61 &   0.79  & 0.999 &$2.3 \times 10^{-6}$\\
\hline
\end{tabular}
\end{center}
\caption{
         Parameter estimates and goodness of fit values for the correlation functions in Figure \ref      
         {fig:mem_simandfits_n10} using the Gamma density (\ref{eq:gammamem}) and truncated Mittag-
         Leffler function (\ref{eq:mlfmem}) as fitting templates.  ${}^{\ast}$ The time scale $\tau_{\epsilon}$ is displayed in units of time for comparison with the exponential relaxation $\tau_{a}$. SSE stands for Sum of the Squared Errors.   }
\label{table:parameters}         
\end{table}
Data for the velocity time series starts being collected after a transient of 1000 time steps, after which time the time series becomes second--order stationary.  Figure \ref{fig:cstationarity} shows that after a very short transient of a few hundred time steps, the estimators become very narrowly bounded and no trend with time is evident.   
\subsection{Estimation of  $\vb$ and $\Db$}\label{ssec:estim_vD}
\begin{figure}
 \begin{center}
 \includegraphics[width=6in]{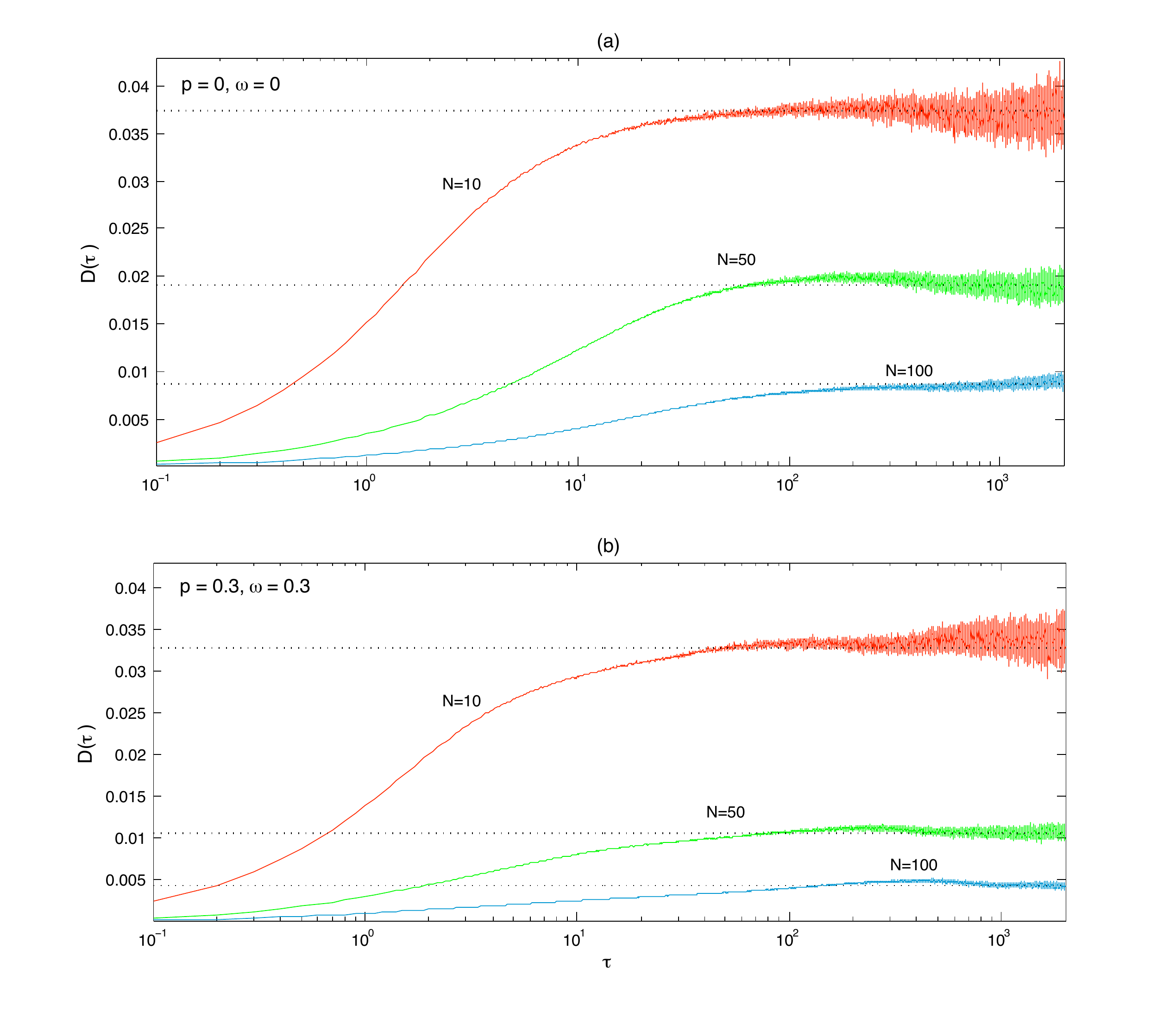}
 \caption{Diffusion coefficients estimated via the Kubo--Green relationship (\ref{eq:DfromGreenKubo}) (dotted black lines), and from the variance of the jump kernel (\ref{eq:estimDtau}) sampled at  various time lags $\tau$ .  Panel (a) shows estimates for purely na\"{i}ve swarms of total population size $N=10$ (red), $N=50$ (green) and $N=100$ (blue).  Panel (b) shows the estimates for informed configurations. The three cases share the same fraction of informed individuals $p=0.3$ and coupling constant $\omega =0.3$, the total population sizes are color coded as in panel (a).  We note that both methods succeed  in providing the asymptotic value of $D$.  There is an overall reduction in diffusivity as the total group size increases.  The diffusivity also decreases in informed groups compared with naive ones of the same total size.}
 \label{fig:Dmem&Djumpkernel}
 \end{center}
\end{figure}

The drift coefficient $v_1$  can be estimated in two ways.  The most straightforward is from the sample mean of the velocity time series $\{v_1,v_2,\ldots,v_T\}$,
\begin{equation}
\hat{v}_1^{\ast} = \frac{1}{T} \sum_{i=1}^{T} v_{i}
\label{eq:meanv1}
\end{equation}
and the other is based on the first moment of the jump kernel,
\begin{equation}
  \label{eq:estimV}
    v_\lambda= \frac{1}{\bar{t}}\int_\re x_1\,\lambda_1(x_1) dx_1 
\end{equation}
where $\bar{t}$ is the mean time between transitions (\ref{eq:tbar}), and $\lambda_1$ is the marginal of the jump kernel $\lambda(\x)$ along the informed coordinate. Since the characteristic time $\bar{t}$ is not known, the estimator requires sampling the jump kernel $\lambda$ at various lags $\tau$.  The characteristic time will be the value of $\tau$ for which the estimator saturates,
\begin{equation}
\hat{\mu}(\tau) =  \frac{1}{T(\tau)}\sum_{i=1}^{T(\tau)}\Delta(x_{1};\tau)
\end{equation}
where $\Delta(x_{1};\tau)$ is the sub--series of position differences along the direction $x_{1}$ sampled at time lag $\tau$ from the \emph{position} time series $\{x_{1},x_{2},\ldots,x_{T}\}$, where $T$ is the total length of the series, and $T(\tau)$ is the length of the sub-series sampled at lag $\tau$.  Of course, the quality of the estimator decreases with $\tau$, because the length of each sub-series is twice as short as the preceding one. The lag dependent drift is then given by
\[
\hat{v}_{\lambda}(\tau)=\frac{\hat{\mu}}{\tau}
\]
and the characteristic time can be calculated as the smallest value of the lag $\tau^{\ast}$ for which the equality
\[
\hat{v}_{\lambda}(\tau^{\ast}) = \hat{v}_{1}^{\ast}
\]
that relates both estimators holds.   Since a parametric form of the memory is already available (see Section \ref{ssec:estim_phi}), the diffusivity can be estimated from the Kubo--Green relationship \cite{kubo91,kenkre03,kenkre09} that relates transport parameters to time correlation functions,
\begin{equation}
 \label{eq:DfromGreenKubo}
 D_{1} = \int_0^\infty \E \left[
                                        \left(
                                          v_{1}(0)   - \bar{v}_{1}
                                        \right)
                                        \left(
                                          v_{1}(\tau) -\bar{v}_{1}
                                        \right)
                                       \right]\,d\tau.
\end{equation}
It can also be estimated from the second moments of the jump kernel
\begin{equation}
  \label{eq:estimD}
 D_1=\frac{1}{2\,\bar{t}}\int_\re (x_1-\mu_1)^2\,\lambda_1(x_1) \,dx_1,
\end{equation}
where an estimator of $D_{1}$ is developed in a similar vein as that of the drift $\hat{v}_{\lambda}$
\begin{equation}
  \label{eq:estimDtau}
  \widehat{D}_1(\tau)=\frac{1}{2\,\tau\,(T(\tau)-1)}\sum_{i=1}^{T(\tau)}\left(\Delta(x_{1};\tau)-\mu(\tau)\right)^{2},
\end{equation}
the diffusion coefficient is the value for which $\hat{D_1}(\tau)$ reaches a plateau.  The behavior of both estimators for the swarm meta--particle is shown in figure \ref{fig:Dmem&Djumpkernel}.  The upper panel shows the results for na\"{i}ve configurations of various total population sizes,  $N=10$ (red), $N=50$ (green) and $N=100$ (blue). The dotted black line corresponds to the estimate of the diffusivity from the velocity time auto--correlation using the Kubo--Green relationship (\ref{eq:DfromGreenKubo}) and the rugged lines of various colors correspond the estimates of the diffusivity based on (\ref{eq:estimDtau}) that vary with the sampling lag $\tau$. The lower panel shows the comparisons between both methods for informed configurations of the same total population sizes as in the upper panel, but including informed individuals for the same coupling constants.  In all the cases the \emph{proportion} of informed individuals $p = N/N_{\rb} =0.3$ was kept constant. We observe that both methods converge to approximately the same value, in both na\"{i}ve and informed configurations.  We note that the characteristic time --the time at which the estimator saturates-- increases with group size.  The width of the oscillations in the estimator (\ref{eq:estimDtau}) increases with the lag $\tau$ due to the finite size of the location time series, since for larger values of $\tau$, the number of data points used in the estimator decreases.
\begin{figure}
 \begin{center}
 \includegraphics[width=6in]{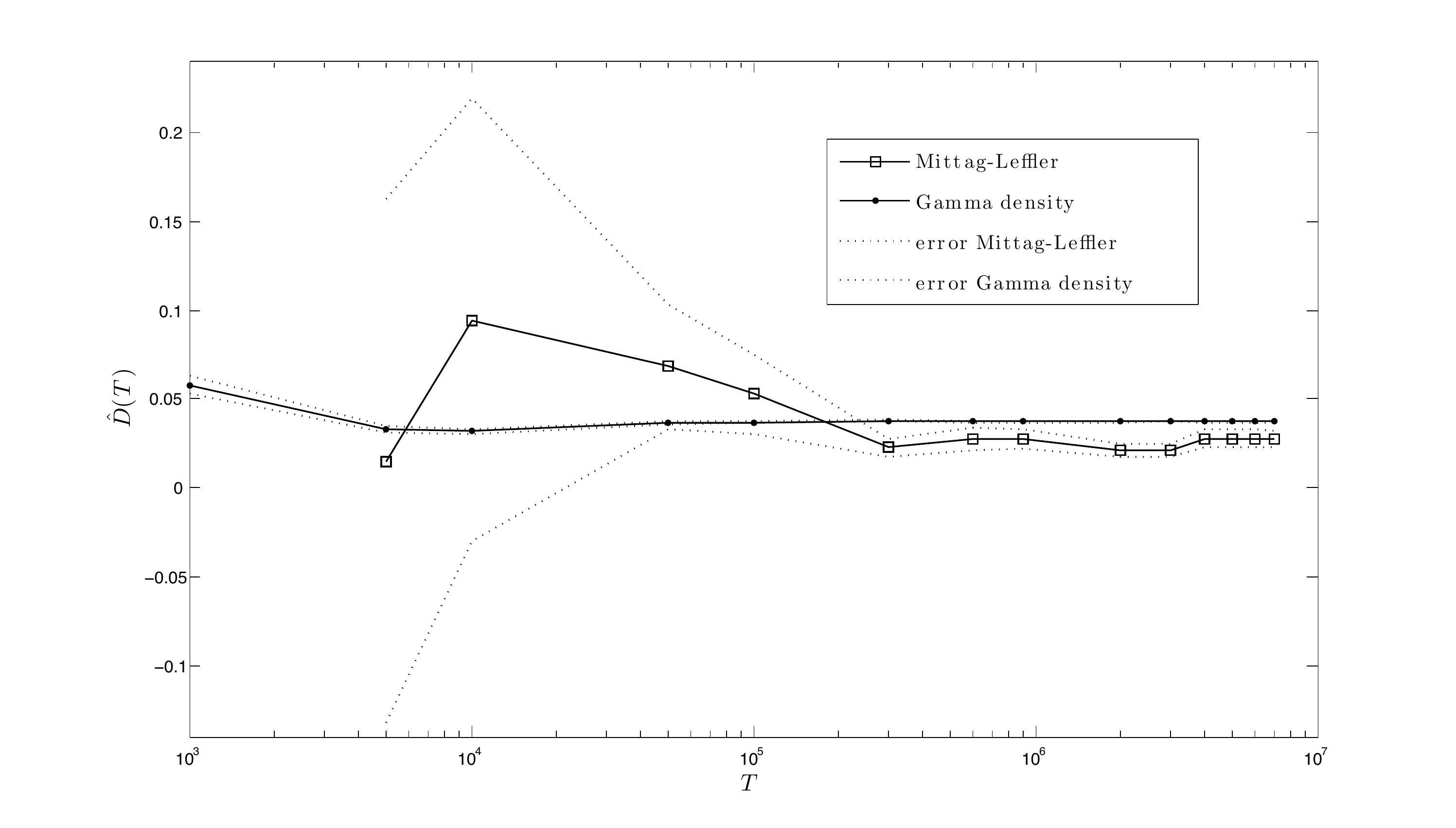}
 \caption{Behavior of the estimator of the diffusion coefficient $\hat{D}(T)$ based on the Kubo--Green relationship (\ref{eq:DfromGreenKubo}) versus the length $T$ of the meta--particle velocity time series. Open squares denote the value of the estimator using a truncated Mittag--Leffler kernel template for the velocity auto--correlation, and black circles correspond to a Gamma density.  In both cases the dotted lines are the 95\% confidence intervals.}
 \label{fig:estimDvsTlogscale}
 \end{center}
\end{figure}
\subsection{Estimation of the time to consensus $\tau_c$}\label{ssec:estim_tauc}
\begin{figure}
 \begin{center}
 \includegraphics[width=6in]{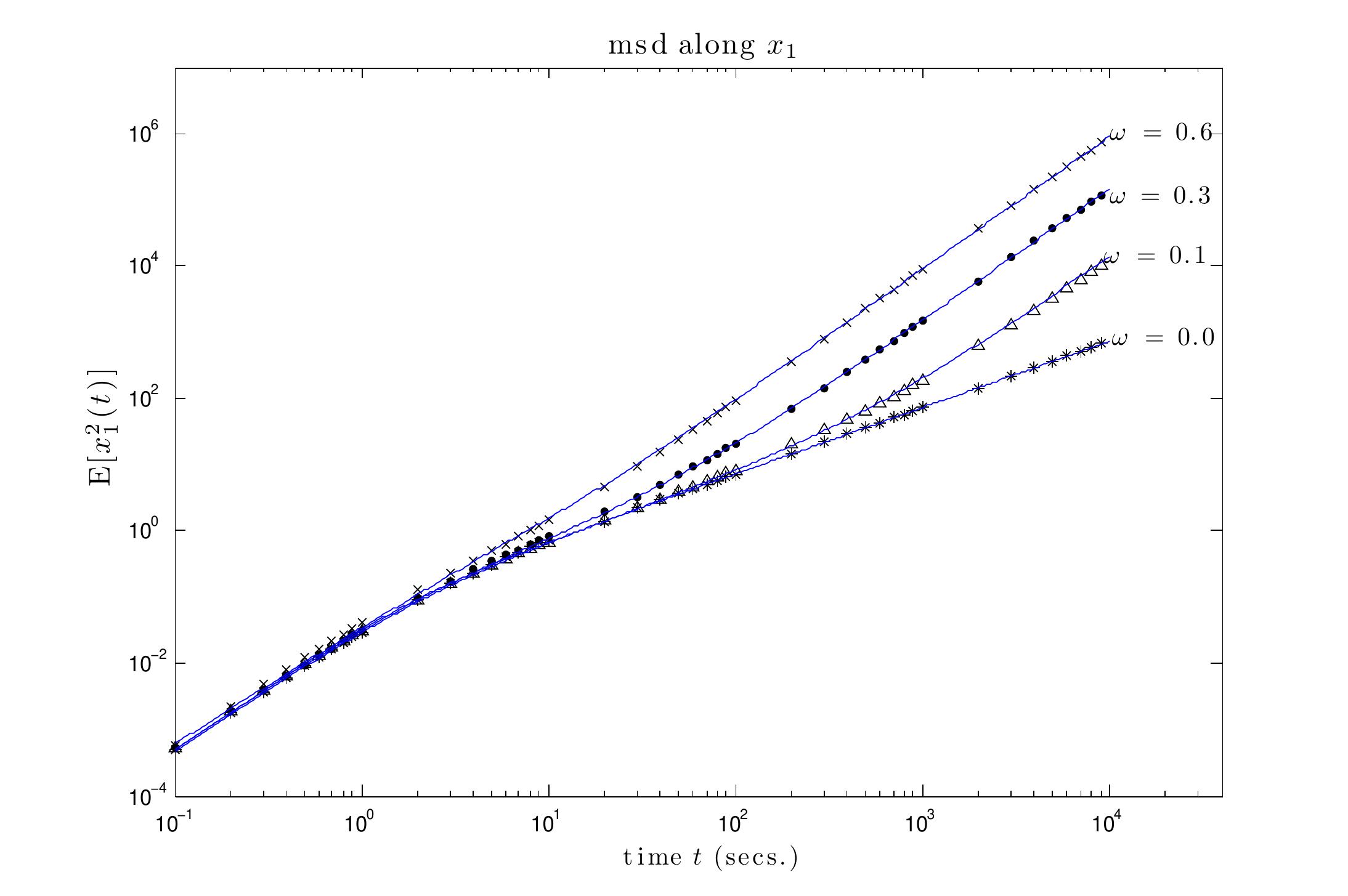}
 \caption{Comparison between the mean squared displacement along the informed direction $x_1$ (\ref{eq:momrb}) estimated from an ensemble of 3000 simulation runs (black marks) and  that obtained from the inverse Laplace transform of the msd (\ref{eq:msd_lpl}) based on the fitted ADEM (blue continuous lines), with parameters estimated from a single simulation run.  We used a Gamma density memory kernel for the lower values of the coupling constant ($\omega=0, \omega = 0.1$) and an exponentially truncated Mittag--Leffler function for the remainder cases ($\omega = 0.3,\omega = 0.6$).  In all cases the total population size consisted of $N=10$ individuals, and informed configurations consisted of one informed individual in all cases.   }
 \label{fig:msd_x1_memFit}
 \end{center}
\end{figure}
We defined crudely the time to consensus $\tau_c$ as the time scale that determines the onset of the quadratic scaling in the mean squared displacement (Figure \ref{fig:msd_x1}) along the informed direction, which in the Laplace domain is given by
\begin{equation}
 \label{eq:lplmsd_1a}
\tilde{m}^{(2)}_1(\epsilon) = \frac{2\,v_1^2}{\epsilon^3}\, \tilde{M}_1^{2}(\epsilon) + \frac{2\,D_1}{\epsilon^2}\, \tilde{M}_1(\epsilon),
\end{equation}
where the coefficients $v_{1}$ and $D_{1}$ can be determined from (\ref{eq:meanv1}) and (\ref{eq:DfromGreenKubo}) respectively. The parameters of the memory are calculated by the method described in Section \ref{ssec:estim_phi}.  Given that the analytical Laplace transforms are known for both memory templates (\ref{eq:gammamem}) and (\ref{eq:mlfmem}), substituting the Laplace transform of the Gamma memory (\ref{eq:gamma_lpl}) into (\ref{eq:lplmsd_1a}) leads to
\begin{equation}
\label{eq:msd_gamma_lpl}
\tilde{m}^{(2)}_{\Gamma}(\epsilon)=\frac{2}{\epsilon^{3}}\left(\tau_{a}^{-1}+\epsilon\right)^{\beta-2}
\left[
     D_{1}\epsilon \left(\tau_{a}^{-1}+\epsilon\right)
     +
     v_{1}^{2}\left(\tau_{a}^{-1}+\epsilon\right)^{\beta}
\right].
\end{equation}
Likewise, for the substituting the Laplace transform of the truncated Mittag--Leffler function (\ref{eq:mlf_lpl}) yields
\begin{equation}
\label{eq:msd_mlf_lpl}
\tilde{m}^{(2)}_{E}(\epsilon)=\frac{2\tau_\epsilon \left(
                                                \tau_{a}^{-1}+\epsilon
                                               \right)^{\alpha-2\beta}
                                }{
                                \epsilon^{3}\left(
                                                 1+\tau_\epsilon \left(
                                                                 \tau_a^{-1} +\epsilon
                                                                  \right)^{\alpha}                   
                                            \right)^{2}
                                }
                                \left(
                                   D_{1}\,\epsilon \left[\tau_{a}^{-1}+\epsilon\right]^{\beta}
                                   +
                                   \tau_{\epsilon} \left[\tau_{a}^{-1}+\epsilon \right]^{\alpha}
                                   \left[
                                   v_{1}^{2}+D_{1}\,\epsilon
                                      \left(
                                      \tau_{a}^{-1}+\epsilon
                                      \right)^{\beta}
                                   \right]
                                \right).
\end{equation}
In the case of the msd for the Gamma density memory (\ref{eq:msd_gamma_lpl}), it is possible to invert analytically the Laplace transform. The msd with the truncated Mittag--Leffler memory can be inverted numerically using the inversion algorithm of de Hoog \cite{deHoog82}.   Before we can use the results of the analytical and numerical inversions of (\ref{eq:msd_gamma_lpl}) and (\ref{eq:msd_mlf_lpl}) we show in Figure \ref{fig:msd_x1_memFit} comparisons between the msd obtained from an ensemble of simulation runs of the swarm meta-particle (black marks) and that obtained by inversion of the Laplace transforms of the mean squared displacements (\ref{eq:msd_gamma_lpl}) and (\ref{eq:msd_mlf_lpl}) (blue lines) based on the ADEM assumption,  with parameters estimated from a single simulation run of the SPP, using the method outlined in Sections \ref{ssec:estim_vD} and \ref{ssec:estim_phi}.  In all cases the method based in the ADEM is able to capture accurately both the transient and the asymptotic behavior. In order to use these results to calculate the time to consensus, $\tau_{c}$ we first note that in the simpler case of a memory of the form of a Dirac distribution $\delta(t)$, Laplace inversion of (\ref{eq:lplmsd_1a}) is straightforward, 
\[
m^{(2)}_1(t) = v_1^2 \,t^2 + 2 D_1\,t,
\]
in which case $\tau_c$ is the smallest time scale for which the contribution due to advection is larger than that of diffusion,
\[
v_1^2 \,t^2>2 D_1\,t,
\]
which leads to
\begin{equation}
\tau_c = \frac{2D_1}{v_1^2}.
\end{equation}
A similar procedure can be carried out for non-trivial choices  for the memory. The first of these is an exponential memory with a relaxation time scale $\tau_{a}=1/b$.   This functional form dominates the asymptotic behavior in both the Gamma and the truncated Mittag--Leffler memory kernels if the anomalous time scale $\tau_{\epsilon}$ in the latter is sufficiently fast compared with $1/b$.  An analogous procedure yields the time to consensus
\begin{equation}
 \label{eq:tauc_exp}
\tau_c \approx \frac{2D_1 \left(1-\frac{1-\exp(-b\,\tau_c )
                                                            }{
                                                            b\,\tau_c
                                                            }
                                          \right)
                               }{
                               v_1^2 \left( 1+ \frac{6}{b^2\,\tau_c^2}
                                                    -  \frac{4}{b\,\tau_c}
                                                    + \left[
                                                           \frac{b\,\tau_c -6}{b^2\tau_c^2}
                                                        \right]\,\exp(-b\,\tau_c)
                                          \right),
                                  }
\end{equation}
which requires an iterative solution. The full Gamma kernel (\ref{eq:gammamem}) results in
\begin{equation}
\label{eq:msdGamma_1}
\tau_c = \frac{2D_1\left(1+ \frac{\beta-1}{b \tau_c}  + \frac{
                                                                                 (b\tau_c)^{-\beta}
                                                                                 }{
                                                                                  \Gamma(1-\beta)
                                                                                 } 
                                                                                 \left[ \exp(-b \tau_c)-(\beta 
                                                                                         +b 
                                                                                         \tau_c-1)\,\mathcal{E}_{\beta}
                                                                                         (b \tau_c)
                                                                                 \right]                      
                              \right)
                    }{
                    v_1^2 \left(
                                       1+ \frac{6+4b\tau_c(\beta-1)+2\beta(2\beta-5)
                                              }{b^2 \tau_c^2
                                              }
                                       +
                                       \frac{(b\tau_c)^{2(1-\beta)}
                                       }{
                                       \Gamma(4-2\beta)}
                                      \left[
                                               \exp(-b\tau_c)(2\beta+b\tau_c-1)
                                                - R(b,\beta,\tau_c)
                                      \right]
                                \right)
                    }
\end{equation}
where
\begin{equation}
R(b,\beta,\tau_c)=  ( 6+b^2\tau_c^2+4b\tau_c(\beta-1))+2\beta(2\beta-5)\,\mathcal{E}_{2\beta-3}(b\tau_c)
\end{equation}
and $\mathcal{E}_{\alpha}(x)$ is the exponential integral
\[
\mathcal{E}_{\alpha}(x) =\int_{1}^{\infty}\frac{e^{-x\,t}}{t^{\alpha}}\,dt.
\]
Unfortunately, we were unable to find an analytical inversion of the Laplace transform of  (\ref{eq:msd_mlf_lpl}) for a  Mittag--Leffler memory kernel. However, both memory kernels are dominated asymptotically by the exponential truncation.  For simplicity, we used the exponential approximation (\ref{eq:tauc_exp}) of the time to consensus for the macroscopic analysis of the efficiency of collective decision making for various group sizes, values of the coupling constants and proportions of informed individuals. 
\subsection{Results}\label{ssec:results}
\begin{figure}
 \begin{center}
 \includegraphics[width=6in]{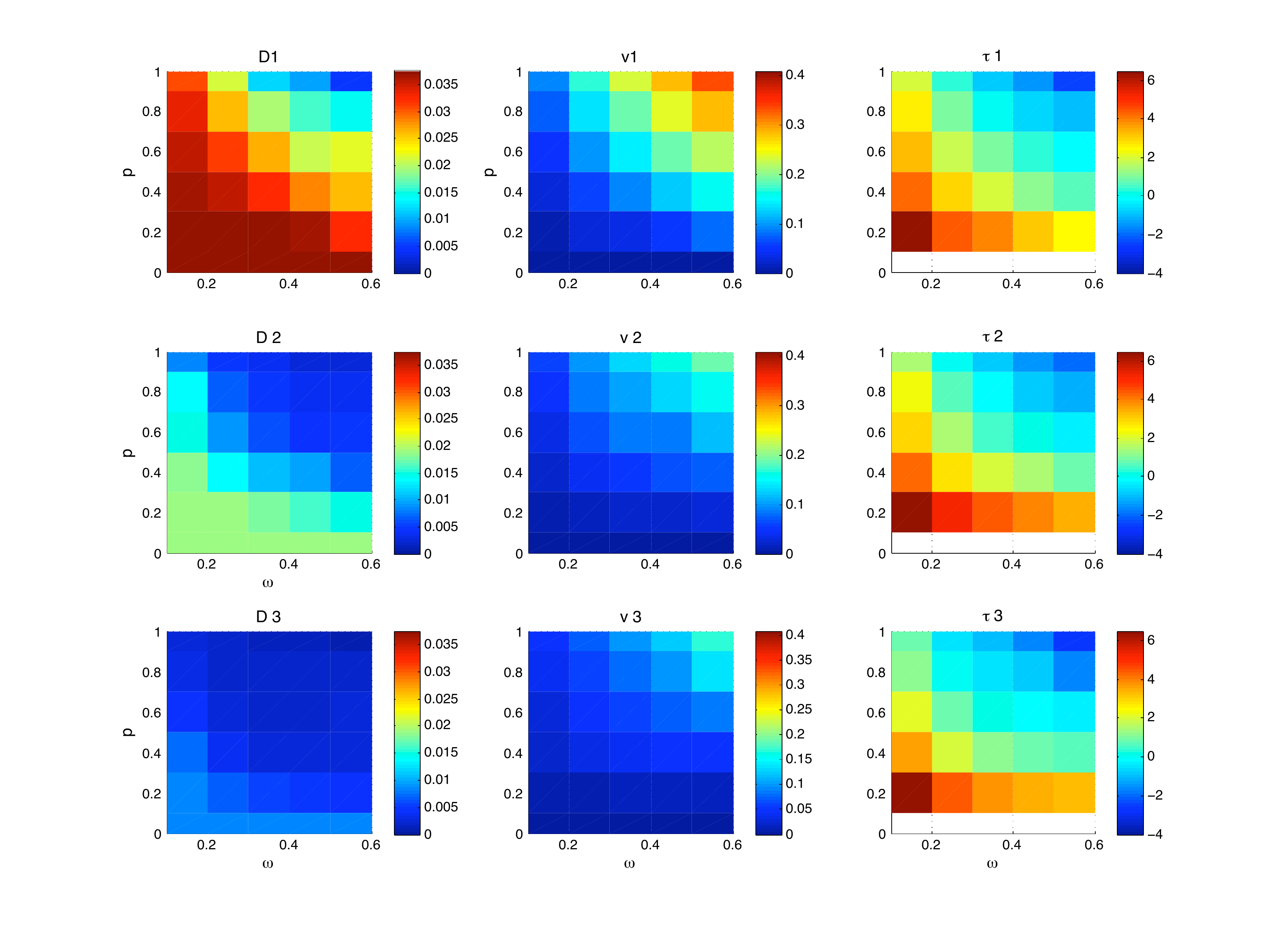}
 \caption{Estimates of the diffusion coefficient (left column) along the informed direction $D$, the mean group speed (center column) $v$, and time to consensus (right) $\tau$ for various values of the proportion of informed individuals $p$ (vertical axis), coupling constant $\omega$ (horizontal axis), and total population sizes. The first row ($D_{1},v_{1}, \tau_{1}$) corresponds to the case $N=10$, the second ($D_{2},v_{2},\tau_{2}$) to $N=50$ and the third ($D_{3},v_{3},\tau_{3}$) to $N=100$.}
 \label{fig:phasediagram}
 \end{center}
\end{figure}

Figure \ref{fig:phasediagram} shows estimates of the three key macroscopic parameters of swarm meta-particles.  The magnitude of the diffusivity $D_{1}$ along the informed direction (left column), the drift $v_{1}$ (center column), and the time to consensus $\tau_{c}$ (right column, logarithmic scale) for three total population sizes $N=10$ (top row), $N=50$ (center row) and $N=100$ (lower row).  In all the graphs the horizontal axis corresponds to the coupling constant $\omega \in [0,0.6]$, and the vertical axis to the relative fraction $p$ of the informed population size to the whole group.   We see that the precision of the collective decision, measured by the ratio of the diffusivities along both coordinates (\ref{eq:precision}) increases with the coupling constant and the number of informed individuals.  Similarly, the degree of consensus (\ref{eq:consensus}), measured by the ratio of the drift $v_{1}$ to the individual particle speed, increases as well with the coupling constant and the informed fraction.  Smaller groups move faster than larger ones, but at the cost of a loss in precision.   Finite size effects are of paramount importance in this class of problems.  Given that the diffusivity decreases with group size as was also detected before \cite{grunbaum08},  traditional approaches where macroscopic quantities are calculated in the limit of very large population sizes are not particularly useful in this context. The time to consensus $\tau_{c}$ decreases with increasing number of informed individuals and coupling strength.  This is not surprising since it is tied to first order to the ratio $D_{1}/v_{1}^{2}$.  Interestingly, it appears to be invariant to group size and controlled by the time scale of the exponential relaxation $\tau_{a}$ which increases as the group size grows.
\section{Final comments}\label{sec:discussion}
This study suggests that both the transient and the asymptotic regimes of swarming populations --with strong alignment  and in the presence of an orientation bias-- can be concisely approximated by an advection--diffusion equation with memory.  The presence of an orientation bias together with macroscopic bursts of alignment, alternating with an unpolarized phase,  lead to quite non-trivial time correlations in the mean group velocity, which persist over macroscopically relevant time scales.  These must be explicitly accounted for in order to capture accurately the macroscopic parameters that typify the various collective states together with their characteristic time scales.  This observation is consistent with recent results by Gr\"{u}nbaum \emph{et al} \cite{grunbaum08} who found that local-in-time advection-diffusion equations even with density dependent coefficients could not fully capture the fluxes of individual-based models of swarming populations when alignment was an important contributor to the dynamics at the level of the individual particle.  That study focused on looking at the fluxes of fission--fusion populations, without informed individuals, for various values of the density in order to try to find a functional form that fitted the dependence of the transport coefficients on the population density.  We explored a much more limited range of population sizes,  but instead looked in more detail at the \emph{temporal} dependence of the mean squared displacement, and the various transport behaviors shown at each time scale.  Of course, both methods are not in opposition but complement each other.  In the future, we would like to integrate both approaches in such a way that both the density--dependence and memory effects are included in a single transport model of swarming populations with alignment. \\
We find that the mean group velocity increases as a power law of the coupling constant, and that the exponent of the power law decreases as the number of informed individuals  increases.  We also find --in agreement with earlier work \cite{couzin05}-- that the total group size has a dramatic impact in the collective transport properties.  Smaller groups tend to move with higher velocities, but at the expense of a higher diffusivity and thus less precise decisions. This may have important implications for evolutionary studies of simple models of collective--decision making, where there is presumably costs associated with recruiting informed individuals into the population, by having a relatively high value of the coupling constant and by making erroneous decisions (Vishwesha Guttal \emph{et al}, personal communication). If some value of the mean group velocity along the informed direction is optimal in a way that maximizes a measure of individual--level fitness, there are a number of possible ways to achieve it.  One possible path is to have a small number of informed individuals, each with a relatively high coupling strength, while another is to have a larger number of informed individuals but with a much smaller coupling strength.  A very rich trade--off space is likely to occur in this class of systems, particularly if one allows for variability in total population size.\\
Remarkably,  the efficiency of collective decision--making, understood as the time scale at which an effective drift  becomes detectable over the diffusive component of the meta--particle random walk, seems to be invariant with respect to group size.  What seems to determine the efficiency is a combination of  the fraction of informed individuals and the strength of the orientation bias.  This arises from the fact that this quantity ultimately depends on the ratio $D/v^{2}$ and the characteristic time scale $\tau_{a}$ of the exponential decay in the memory  (\ref{eq:tauc_exp}). \\
%
 The time velocity auto--correlation emerges from the ADEM approach as the key macroscopic summary statistic. It quantifies the relative contributions to macroscopic transport from each collective behavior, and allows the specification of their characteristic time scales.  Although the ability of time correlation functions to connect microscopic dynamics with observed macroscopic regimes has been known in non--equilibrium statistical physics for at least four decades since the seminal work of Kubo \cite{kubo57}, Mori \cite{mori58}, Green \cite{green60}, Zwanzig \cite{zwanzig65}, Montroll \cite{montroll73} and Kenkre \cite{kenkre77}, to our knowledge it is a relatively unexplored concept in movement and spatial ecology, where Markovian models have dominated the scene \cite{okubo01}, perhaps with the notable exception of correlated random walks \cite{codling08,getz08,nathan08}.  We would like to emphasize a subtle point though, which is that the temporal memory of the ADEM does not necessarily imply that the individual walker has information about the past in order to make movement decisions about the future. The memory arises naturally as a result of the ensemble average of a continuous time random walk in the presence of a wide range of transition rates.  These can result from internal properties --like an updating clock with a `fat tail' instead of an exponential one-- or external factors such as behavioral variability due to complicated social interactions or spatial structure in the landscape that results in slip/stick dynamics;  these can occur quite naturally if there are corridors with preferential directions of motion alternating with regions where movement can be described with Brownian motion.   We believe that this ecological interpretation of the time velocity auto--correlation function is likely to be useful not only to  unravel the connections between individual--based models of movement and dispersal and their continuum approximations as we have seen in this study, but also for other areas of ecology where interdependencies between an individual organism's dispersal strategy,  spatial heterogeneity in the landscape, and temporal variability in resource availability become intertwined in observed individual trajectories, particularly in the nascent field of movement ecology. \\
Future work will be devoted to a generalization of the SPP model to density-dependent asynchronous updating, in the sense that each of the social interactions is associated with an exponential clock that is parameterized by the local density, in a similar way to what is done in locally regulated models of plant population dynamics with spatial structure \cite{bol97,bol00}.   Given that the CTRW-ADEM can predict the the full density and not just the first moments, future work will be devoted to this issue in order to explore first passage times.  We will also explore the situation when there are two conflicting preferential directions, where it remains to be seen whether the ADEM has the capability of  capturing the bifurcations that have been detected in individual--based simulations \cite{couzin05}.  This will require generalizations of the ADEM involving anisotropy in the memory kernel. 
\section{Acknowledgements} \label{sec:acknowledgements}
The authors are grateful for the support received from the National Science Foundation (Award ID:EF-0434319) and DARPA (Award ID:HR001-05-1-0057).  Insightful discussions with I. D.  Couzin, L.Giuggioli, V.M. Kenkre and F.Bartumeus are gratefully acknowledged.\\
\bibliographystyle{plain}

\begin{thebibliography}{10}

\bibitem{aoki82}
I.~Aoki.
\newblock A simulation study on the schooling mechanism in fish.
\newblock {\em Bulletin of the Japanese Society of Scientific Fisheries},
  48(8):1081--1088, 1982.

\bibitem{aranson07}
Igor~S. Aranson, Andrey Sokolov, John~O. Kessler, and Raymond~E. Goldstein.
\newblock Model for dynamical coherence in thin films of self propelled
  microorganisms.
\newblock {\em Physical Review E}, 75:040901{(R)}, 2007.

\bibitem{berkowitz06}
Brian Berkowitz, Andrea Cortis, Marco Dentz, and Harvey Scher.
\newblock Modeling non--{F}ickian transport in geological formations as a
  continuous time random walk.
\newblock {\em Reviews of Geophysics}, 44(2):Art. No. RG2003/2006, 2006.

\bibitem{bertin06}
Eric Bertin, Michel Droz, and Guillaume Gregoire.
\newblock Boltzmann and hydrodynamic description for self-propelled particles.
\newblock {\em Physical Review E (Statistical, Nonlinear, and Soft Matter
  Physics)}, 74(2):022101, 2006.

\bibitem{bol97}
Benjamin Bolker and Stephen~W. Pacala.
\newblock Using moment equations to understand stochastically driven spatial
  pattern formation in ecological systems.
\newblock {\em Theoretical Population Biology}, 52:179--197, 1997.

\bibitem{bol00}
Benjamin Bolker, Pacala~Stephen W., and Simon~A. Levin.
\newblock Moment methods for ecological processes in continuous space.
\newblock In Ulf Dieckmann, Richard Law, and Johan~A.J Metz, editors, {\em The
  Geometry of Ecological Interactions}, volume~1 of {\em Cambridge Studies in
  Adaptive Dynamics}, pages 388--411. Cambridge University Press, 2000.

\bibitem{buhl06}
J~Buhl, DJT Sumpter, ID~Couzin, JJ~Hale, E~Despland, ER~Miller, and SJ~Simpson.
\newblock From disorder to order in marching locusts.
\newblock {\em Science}, {312}({5778}):{1402--1406}, {JUN 2} {2006}.

\bibitem{cavagna06}
Andrea Cavagna, Alessio Cimarelli, Irene Giardina, Alberto Orlandi, Giorgio
  Parisi, Andrea Procaccini, Raffaele Santagati, and Fabio Stefanini.
\newblock {New statistical tools for analyzing the structure of animal groups}.
\newblock {\em Mathematical Biosciences}, {214}({1-2}):{32--37}, {JUL-AUG}
  {2008}.

\bibitem{chuang07}
Yao-Li Chuang, Maria~R. D'{O}rsogna, Daniel Marthaler, Andrea~L. Bertozzi, and
  Lincoln~S. Chayes.
\newblock State transitions and the continuum limit for a 2{D} interacting,
  self-propelled particle system.
\newblock {\em Physica D}, 232:33--47, 2007.

\bibitem{codling08}
Edward~A. Codling, Michael~J. Plank, and Simon Benhamou.
\newblock {Random walk models in biology}.
\newblock {\em Journal of the Royal Society Interface}, {5}({25}):{813--834},
  {AUG 6} {2008}.

\bibitem{coifman08}
R.~R. Coifman, I.G. Kevrekidis, S.~Lafon, M.~Maggioni, and B.~Nadler.
\newblock Diffusion maps, reduction coordinates, and low dimensional
  representation of stochastic systems.
\newblock {\em Multiscale Modeling \& Simulation}, {7}({2}):{842--864}, {2008}.

\bibitem{conradt09}
Larissa Conradt and Christian List.
\newblock Group decisions in humans and animals: a survey introduction.
\newblock {\em Philosophical Transactions of the Royal Society {B}-Biological
  Sciences}, {364}({1518}):{719--742}, {MAR 27} {2009}.

\bibitem{conradt03}
Larissa Conradt and Timothy~J. Roper.
\newblock Group decision--making in animals.
\newblock {\em Nature}, 421(8):155--158, 2003.

\bibitem{cortis04}
Andrea Cortis, Claudio Gallo, Harvey Scher, and Brian Berkowitz.
\newblock Numerical simulation of non-fickian transport in geological
  formations with multiple scale heterogeneities.
\newblock {\em Water Resources Research}, 40:W04209, 2004.

\bibitem{couzin05}
Ian~D. Couzin, Jens Krause, Nigel~R. Franks, and Simon~A. Levin.
\newblock Effective leadership and decision-making in animal groups on the
  move.
\newblock {\em Nature}, 433:513--516, 2005.

\bibitem{couzin02}
Ian~D. Couzin, Jens Krause, Richard James, Graeme~D. Ruxton, and Nigel~R.
  Franks.
\newblock Collective memory and spatial sorting in animal groups.
\newblock {\em Journal of Theoretical Biology}, 218:1--11, 2002.

\bibitem{czirok99}
Andr\'{a}s Czir\'{o}k, Albert-L\'{a}szl\'{o} Barab\'{a}si, and Tam\'{a}s
  Vicsek.
\newblock Collective motion of self-propelled particles: Kinetic phase
  transition in one dimension.
\newblock {\em Physical Review Letters}, 82(1):209--212, 1999.

\bibitem{deHoog82}
F.R. de~Hoog, J.H. Knight, and A.N Stokes.
\newblock An improved method for numerical inversion of {L}aplace transforms.
\newblock {\em SIAM Journal on Scientific and Statistical Computing},
  3(3):357--366, 1982.

\bibitem{eftimie07}
R.~Eftimie, G.~de~Vries, and M.~A. Lewis.
\newblock Complex spatial group patterns result from different animal
  communication mechanisms.
\newblock {\em Proceedings of the National Academy of Sciences of the United
  States of America}, {104}({17}):{6974--6979}, {APR 24} {2007}.

\bibitem{rad05}
Radek Erban, Ioannis~G. Kevrekidis, and Hans~G. Othmer.
\newblock An equation-free computational approach for extracting
  population-level behavior from individual-based models of biological
  dispersal.
\newblock {\em Physica D-- Nonlinear phenomena}, {215}({1}):{1--24}, {MAR 1}
  2006.

\bibitem{flierl99}
G~Flierl, D~Gr\"{u}nbaum, S~Levin, and D~Olson.
\newblock From individuals to aggregations: The interplay between behavior and
  physics.
\newblock {\em Journal of Theoretical Biology}, {196}({4}):{397--454}, {FEB 21}
  1999.

\bibitem{gar85}
C.W. Gardiner.
\newblock {\em Handbook of Stochastic Methods}.
\newblock Springer-Verlag, second edition, 1985.

\bibitem{getz08}
Wayne~M. Getz and David Saltz.
\newblock {A framework for generating and analyzing movement paths on
  ecological landscapes}.
\newblock {\em Proceedings of the National Academy of Sciences of the United
  States of America}, {105}({49}):{19066--19071}, {DEC 9} {2008}.

\bibitem{green60}
MS~Green.
\newblock Comment on a paper of mori on time-correlation expressions for
  transport properties.
\newblock {\em Physical Review}, {119}({3}):{829--830}, {1960}.

\bibitem{grunbaum08}
Daniel Gr\"{u}nbaum, Karen Chan, Elizabeth Tobin, and Michael~T. Nishizaki.
\newblock Non--linear advection-diffusion equations approximate swarming but
  not schooling populations.
\newblock {\em Mathematical Biosciences}, 214(1-2):38--48, {JUL-AUG} 2008.

\bibitem{grunbaum94}
Daniel Gr\"{u}nbaum and Akira Okubo.
\newblock Modeling social animal aggregations.
\newblock In Simon~A. Levin, editor, {\em Frontiers in Theoretical Biology},
  volume 100 of {\em Lecture Notes in Biomathematics}, pages 296--325.
  Springer-Verlag, 1994.

\bibitem{halmos46}
Paul~R. Halmos.
\newblock The theory of unbiased estimation.
\newblock {\em Annals of Mathematical Statistics}, 17(1):34--43, 1946.

\bibitem{haus87}
JW~Haus and KW~Kehr.
\newblock Generalized effective--medium approximation for particle --transport.
\newblock {\em Physical Review B}, {36}({10}):{5639--5642}, {OCT 1} {1987}.

\bibitem{huth92}
A.~Huth and C.~Wissel.
\newblock The simulation of the movement of fish schools.
\newblock {\em Journal of Theoretical Biology}, 156:365--385, 1992.

\bibitem{kenkre77}
V.~M. Kenkre.
\newblock The generalized master equation and its applications.
\newblock In Uzi Landman, editor, {\em Statistical Mechanics and statistical
  methods in theory and application}, pages 441--461. Plenum Publishing
  Corporation, 1977.

\bibitem{kenkre09}
V.~M. Kenkre, Z.~Kalay, and P.~E. Parris.
\newblock Extensions of effective-medium theory of transport in disordered
  systems.
\newblock {\em Physical Review {E}}, {79}({1, Part 1}), {JAN} {2009}.

\bibitem{kenkre03}
V.M. Kenkre.
\newblock Memory formalism, nonlinear techniques and kinetic equation
  approaches.
\newblock In V.M. Kenkre and K.~Lindenberg, editors, {\em Modern Challenges in
  Statistical Mechanics: Patterns, Noise and the Interplay of Nonlinearity and
  Complexity}, pages 63--102. American Institute of Physics, 2003.

\bibitem{kev03}
Ioannis~G. Kevrekidis, C.~William Gear, James~M. Hyman, Panagiotis~G.
  Kevrekidis, Olof Runborg, and Constantinos Theodoropulos.
\newblock Equation--free, coarse--grained multiscale computation:enabling
  microscopic simulators to perform system--level analysis.
\newblock {\em Communications in Mathematical Sciences}, 1(4):715--762, 2003.

\bibitem{klafter80}
J~Klafter and R~Silbey.
\newblock Derivation of the continuous time random walk equation.
\newblock {\em Physical Review Letters}, 44(2):55--58, 1980.

\bibitem{kolpas07}
Allison Kolpas, Jeff Moehlis, and Ioannis~G. Kevrekidis.
\newblock Coarse--grained analysis of stochasticity--induced switching between
  collective motion states.
\newblock {\em Proceedings of the National Academy of Sciences},
  104(14):5931--5935, 2007.

\bibitem{kubo57}
R~Kubo.
\newblock Statistical-mechanical theory of irreversible processes .1. general
  theory and simple applications to magnetic and conduction problems.
\newblock {\em Journal of the Physical Society of Japan}, {12}({6}):{570--586},
  {1957}.

\bibitem{kubo91}
Ryogo Kubo, Morikazu Toda, , and Natsuki Hashitsume.
\newblock {\em Statistical physics II: Nonequilibrium Statistical Mechanics}.
\newblock Springer-Verlag, 1991.

\bibitem{merrifield06}
A.~Merrifield, Mary~R. Myerscough, and N.~Weber.
\newblock Statistical tests for analysing directed movement of self-organising
  animal groups.
\newblock {\em Mathematical Biosciences}, {203}({1}):{64--78}, {SEP} {2006}.

\bibitem{metzler04}
R~Metzler and J~Klafter.
\newblock {The restaurant at the end of the random walk: recent developments in
  the description of anomalous transport by fractional dynamics}.
\newblock {\em Journal of Physics {A}-Mathematical and General},
  37(31):{R161--R208}, AUG 6 2004.

\bibitem{mogilner99}
A~Mogilner and L~Edelstein-Keshet.
\newblock A non-local model for a swarm.
\newblock {\em Journal of Mathematical Biology}, {38}({6}):{534--570}, {JUN}
  {1999}.

\bibitem{montroll73}
E.W. Montroll and H.~Scher.
\newblock Random walks on lattices iv. continuous time random walks and
  influence of absorbing boundaries.
\newblock {\em Journal of Statistical Physics}, 9(2), 1973.

\bibitem{moon07}
Sung~Joon Moon, B.~Nabet, Naomi~E. Leonard, Simon~A. Levin, and I.~G.
  Kevrekidis.
\newblock Heterogeneous animal group models and their group-level alignment
  dynamics: An equation-free approach.
\newblock {\em Journal of Theoretical Biology}, {246}({1}):{100--112}, {MAY 7}
  {2007}.

\bibitem{mori58}
H~Mori.
\newblock Statistical-mechanical theory of transport in fluids.
\newblock {\em Physical Review}, {112}({6}):{1829--1842}, {1958}.

\bibitem{nabet09}
Benjamin Nabet, Naomi~E. Leonard, Iain~D. Couzin, and Simon~A. Levin.
\newblock {Dynamics of Decision Making in Animal Group Motion}.
\newblock {\em Journal of Nonlinear Science}, {19}({4}):{399--435}, {AUG}
  {2009}.

\bibitem{nathan08}
Ran Nathan, Wayne~M. Getz, Eloy Revilla, Marcel Holyoak, Ronen Kadmon, David
  Saltz, and Peter~E. Smouse.
\newblock {A movement ecology paradigm for unifying organismal movement
  research}.
\newblock {\em Proceedings of the National Academy of Sciences of the United
  States of America}, {105}({49}):{19052--19059}, {DEC 9} {2008}.

\bibitem{okubo01}
Akira Okubo and Simon Levin.
\newblock {\em Difussion and Ecological Problems: Modern Perspectives},
  volume~14 of {\em Interdisciplinary Applied Mathematics}.
\newblock Springer-Verlag, second edition, 2001.

\bibitem{orfanidis96}
S.J. Orfanidis.
\newblock {\em Optimum Signal Processing. An Introduction}.
\newblock Prentice-Hall, second edition, 1996.

\bibitem{parrish02}
Julia~K. Parrish, Steven~V. Viscido, and Daniel G\"{u}nbaum.
\newblock Self-organized fish schools: An examination of emergent properties.
\newblock {\em Biological Bulletin}, 202:296--305, 2002.

\bibitem{podlubny99}
Igor Podlubny.
\newblock {\em Fractional Differential Equations}.
\newblock Mathematics in Science and Engineering. Academic Press, 1999.

\bibitem{reynolds87}
Craig~W. Reynolds.
\newblock Flocks, herds and schools: A distributed behavioral model.
\newblock {\em Computer Graphics}, 21(4):25--34, 1987.

\bibitem{sokolov07}
Andrey Sokolov, Igor~S. Aranson, John~O. Kessler, and Raymond~E. Goldstein.
\newblock Concentration dependence of the collective dynamics of swimming
  bacteria.
\newblock {\em Physical Review Letters}, 98:15802, 2007.

\bibitem{sumpter06}
David J.~T. Sumpter.
\newblock The principles of collective animal behavior.
\newblock {\em Philosophical transactions of the royal society B : Biological
  Sciences}, 361:5--22, 2005.

\bibitem{sumpter08}
David J.~T. Sumpter, Jens Krause, Richard James, Iain~D. Couzin, and Ashley
  J.~W. Ward.
\newblock Consensus decision making by fish.
\newblock {\em Current Biology}, {18}({22}):{1773--1777}, {NOV 25} {2008}.

\bibitem{toner05}
John Toner, Yuhai Tu, and Sriram Ramaswamy.
\newblock Hydrodynamics and phases of flocks.
\newblock {\em Annals of Physics}, 318(1):170--244, 2005.

\bibitem{topaz06}
Chad~M. Topaz, Andrea~L. Bertozzi, and Mark~A. Lewis.
\newblock A nonlocal continuum model for biological aggregation.
\newblock {\em Bulletin of Mathematical Biology}, {68}({7}):{1601--1623}, {OCT}
  {2006}.

\bibitem{van01}
N.G. Van~Kampen.
\newblock {\em Stochastic Processes in Physics and Chemistry}.
\newblock North-Holland, 2001.

\bibitem{ward08}
Ashley J.~W. Ward, David J.~T. Sumpter, Lain~D. Couzin, Paul J.~B. Hart, and
  Jens Krause.
\newblock Quorum decision-making facilitates information transfer in fish
  shoals.
\newblock {\em Proceedings of the National Academy of Sciences of the United
  States of America}, {105}({19}):{6948--6953}, {MAY 13} {2008}.

\bibitem{west97}
B.J. West, P.~Grigolini, R.~Metzler, and T.F. Nonnenmacher.
\newblock Fractional diffusion and levy stable processes.
\newblock {\em Physical Review {E}}, 55({1, Part A}):99--106, JAN 1997.

\bibitem{west03}
Bruce~J. West, Mauro Bologna, and Paolo Grigolini.
\newblock {\em Physics of fractal operators}.
\newblock Springer, 2003.

\bibitem{yates09}
Christian~A. Yates, Radek Erban, Carlos Escudero, Iain~D. Couzin, Jerome Buhl,
  Ioannis~G. Kevrekidis, Philip~K. Maini, and David~J.T. Sumpter.
\newblock United by noise: Randomness helps swarms stay together.
\newblock {\em Proceedings of the National Academy of Sciences},
  106(14):5464--5469, 2009.

\bibitem{zwanzig65}
Robert Zwanzig.
\newblock Time-correlation functions and transport coefficients in statistical
  mechanics.
\newblock {\em Annual Review of Physical Chemistry}, {16}:{67--\&}, {1965}.

\end{thebibliography}

\end{document}